\documentclass[10pt, preprint]{aastex}
\pdfoutput=1

\usepackage{epsfig}
\usepackage{amsmath}
\usepackage{upgreek}
\usepackage{natbib}
\usepackage{graphicx}
\usepackage{longtable}

\begin{document}

\title{The abundance of H$_2$O and HDO in Orion KL from Herschel/HIFI\thanks{{\it Herschel} is an ESA space observatory with science instruments provided by European-led Principal Investigator consortia and with important participation from NASA.}}

\author{Justin L. Neill, Shiya Wang, Edwin A. Bergin, Nathan R. Crockett, C\'{e}cile Favre}
\affil{Department of Astronomy, University of Michigan, 500 Church Street, Ann Arbor, MI 48109, USA; jneill@umich.edu}

\author{Ren\'{e} Plume}
\affil{Department of Physics \& Astronomy and the Institute for Space Imaging Sciences, University of Calgary, Calgary, AB T2N IN4, Canada}

\author{Gary J. Melnick}
\affil{Harvard-Smithsonian Center for Astrophysics, 60 Garden Street, MS 66, Cambridge, MA 02138, USA}

\begin{abstract}
Using a broadband, high spectral resolution survey toward Orion KL acquired with Herschel/HIFI as part of the HEXOS key program, we derive the abundances of H$_2$O and HDO in the different spatial/velocity components associated with this massive star-forming region: the Hot Core, Compact Ridge, and Plateau.  A total of 20 transitions of H$_2$$^{18}$O, 14 of H$_2$$^{17}$O, 37 of HD$^{16}$O, 6 of HD$^{18}$O, and 6 of D$_2$O are used in the analysis, spanning from ground state transitions to over 1200 K in upper-state energy.  Low-excitation lines are detected in multiple components, but the highest-excitation lines ($E_u >$ 500 K) are well modeled as emitting from a small ($\sim 2''$) clump with a high abundance of H$_2$O ($\chi = 6.5 \times 10^{-4}$ relative to H$_2$) and a HDO/H$_2$O ratio of 0.003.  Using high spatial resolution ($1.5'' \times 1.1''$) images of two transitions of HDO measured by ALMA as part of its science verification phase, we identify this component as located near, but not directly coincident with, known continuum sources in the Hot Core region.  Significant HDO/H$_2$O fractionation is also seen in the Compact Ridge and Plateau components.  The outflowing gas, observed with both emission and absorption components, has a lower HDO/H$_2$O ratio than the compact components in Orion KL, which we propose could be due to modification by gas-phase shock chemistry.
\end{abstract}

\keywords{ISM: abundances -- astrochemistry -- ISM: molecules  -- ISM: individual (Orion KL)}

\section{Introduction}

Water is a central molecule in the physics and chemistry of the interstellar medium \citep{vandishoeck11, bergin12, melnick09, caselli12}.   In the cold, dense stages of star formation, water is often the dominant constitutent of the ice mantles that harbor most of the heavy atoms \citep{gibb04, oberg11} and therefore plays a significant role in the formation of the rich organic molecular inventory that is believed to result largely from chemistry on grain surfaces \citep{herbst09}.  In warmer regions ($T > 100$ K), the ice mantles evaporate and water can be one of the major gas phase constituents behind molecular hydrogen.  Finally, at very high temperatures ($T > 400$ K), gas-phase reactions of atomic oxygen with H$_2$ can convert all oxygen not in CO into water on fast timescales \citep{kaufman96, bergin98}.  Due to its high dipole moment, gas-phase water can be detected through strong transitions in the submillimeter and infrared that are important in the energy balance of the molecular cloud \citep{neufeld95}.  The HDO/H$_2$O abundance ratio is also a powerful diagnostic of the evolution of star-forming regions, due to the strong sensitivity of deuterium fractionation processes to physical conditions, particularly temperature \citep{millar03}.  This ratio is posited as a tracer of the possible link between interstellar and cometary water, holding implications for understanding the mechanism for the delivery of water to the young Earth \citep{bockelee-morvan98, hartogh11, caselli12}.  Furthermore, observations suggest that the chemistry leading to the deuteration of H$_2$O is very different from that of other organic molecules such as HCN, H$_2$CO, and CH$_3$OH \citep{vandishoeck11}.

The Orion Kleinmann-Low nebula (Orion KL) is the nearest massive star-forming region, at a distance of 414 $\pm$ 7 pc \citep{menten07}, with very strong gas-phase water emission in the submillimeter and infrared.  Studies of gas-phase H$_2$O from ground-based observatories are limited by atmospheric absorption of the most emissive transitions at the typical temperatures of molecular clouds.  Most of the exceptions are lines that exhibit maser activity in Orion KL \citep{genzel81, menten90, cernicharo90, cernicharo94, cernicharo99, hirota12}, and so have limited usefulness in characterizing the bulk water abundance.  Therefore, water has been a key focus of space-based observatories in the far-infrared.  The ground state \emph{ortho} transitions ($1_{10}-1_{01}$) of H$_2$O and its isotopologues were measured with a few arcminute beam by the Submillimeter Wave Astonomy Satellite (SWAS) \citep{melnick00} and the Odin satellite \citep{persson07}, and a large number of water lines, both pure rotational and vibration-rotation, were observed by the Infrared Space Observatory (ISO) \citep{vandishoeck98, harwit98, lerate06, cernicharo06}.  HDO toward Orion KL has been characterized from ground-based observatories \citep{turner75, petuchowski88, jacq90, pardo01}, with observations suggesting that Orion KL contains warm gas with significant water deuterium fractionation (that is, [HDO]/[H$_2$O] $\gg$ [D]/[H] $\sim 10^{-5}$).

The Herschel Space Observatory \citep{pilbratt10} enables the most comprehensive studies to date of pure rotational transitions of water in star-forming regions, from ground state transitions to highly excited lines, due to its broad spectral coverage, high spatial resolution as compared to previous space-based observatories ($40''$--$10''$ for the Heterodyne Instrument for the Far Infrared (HIFI) \citep{degraauw10}), and high spectral resolution ($\le$ 1.1 MHz for HIFI, or 0.7--0.2 km s$^{-1}$).  As part of the Herschel Observations of EXtra-Ordinary Sources (HEXOS) key program \citep{bergin10}, a full spectral survey of Orion KL with HIFI (covering the frequency ranges 479.5--1280.0 and 1426.0--1906.8 GHz) has been obtained, in which a number of lines of H$_2$O and its rarer isotopologues (H$_2$$^{18}$O, H$_2$$^{17}$O, HDO, HD$^{18}$O, and D$_2$O) have been detected \citep{melnick10, bergin10, crockett10}.  In this report, we use these transitions to derive the abundances of H$_2$O, HDO, and D$_2$O in the spatial components located within the Herschel beam.

In \S 2, we present details of the HIFI observations, along with Atacama Large Millimeter/Submillimeter Array (ALMA) science verification measurements of two transitions of HDO in Orion KL in the 213--245 GHz spectral region.  This is followed by a description of the methods by which the column densities of H$_2$O and HDO in each spatial component are derived, in \S 3.  These results are summarized and discussed in \S 4, with a focus on the differences in D/H ratios and water abundances between components, and \S 5 concludes.

\section{Observations}

\subsection{HIFI}

Results from the HIFI Orion KL spectral survey have been presented elsewhere \citep{bergin10, crockett10}.  All spectra were acquired between March 2010 and April 2011.  For bands 1-5 (480--1280 GHz), the pointing center of the observations was $\alpha_{\textnormal{J2000}} = 05^\textnormal{h} 35^\textnormal{m} 14^\textnormal{s}.3$, $\delta_{\textnormal{J2000}} = -05^\circ 22'33.7''$, located between the two primary regions of compact molecular emission in the Orion KL region, the Hot Core and Compact Ridge.  For bands 6-7 (1426--1535 and 1573--1906 GHz), due to the smaller HIFI beam, spectra were obtained with two separate pointings, centered on the Hot Core and Compact Ridge, with the Compact Ridge pointing lying $8''$ to the southwest of the Hot Core (see \S 3.5 for further discussion of the two pointings).  In this analysis we have used spectra with the Hot Core pointing, with $\alpha_{\textnormal{J2000}} = 05^\textnormal{h} 35^\textnormal{m} 14^\textnormal{s}.5$, $\delta_{\textnormal{J2000}} = -05^\circ 22'30.9''$.  The half-power beamwidth of Herschel is approximately given by $\theta ('') = 21200/\nu_{\textnormal{GHz}}$.  Because HIFI is a double-sideband spectrometer, the spectra were acquired with a redundancy of 6 for bands 1--5, and redundancy 4 for bands 6--7.  The redundancy is defined as the number of local oscillator settings for which each frequency channel is measured; see \cite{bergin10} for more information on the observation strategy and deconvolution procedure.  The wide band spectrometer was used, which has a spectral resolution of 1.1 MHz.  The spectra were acquired in dual beam switch mode with reference beams lying $3'$ to  the east or west of the science target.  All data presented here were processed with HIPE \citep{ott10} version 5.0, using the standard HIFI deconvolution (\emph{doDeconvolution} task), with the H and V polarizations averaged together in the final data product to improve the signal to noise ratio.  For bands 1--5, because the Herschel beam is larger than the sources of compact molecular emission in Orion KL, calibration was performed using aperture efficiencies.  For bands 6--7, where the Herschel beam is comparable in size to the source of the Hot Core and Compact Ridge regions, main beam efficiencies were used because they better describe the coupling to an extended source.  HIFI aperture and main beam efficiences can be found in \cite{roelfsema12}.  We assume a 10\% calibration uncertainty in all measured line fluxes.  In the figures presented here, all intensities are labeled as main beam brightness temperature ($T_\textnormal{mb}$) for simplicity.  The data in all figures been smoothed to a spectral resolution of approximately 0.7 km s$^{-1}$.

Line identifications for both water and other species were made using the XCLASS program\footnote{https://www.astro.uni-koeln.de/projects/schilke/XCLASS}, which provides the functionality of the CLASS software\footnote{http://www.iram.fr/IRAMFR/GILDAS} along with access to the CDMS and JPL catalogs \citep{muller01, muller05, pickett98}.  The line frequencies, strengths, and lower-state energies presented here come from the fits presented in the catalogs, which draw on spectroscopic data from \cite{delucia72, delucia75, johns85, steenbeckeliers71, steenbeckeliers73, messer84, lovas78, bellet70, benedict70}.  The molecular dipole moment comes from \cite{dyke73}. A comprehensive analysis of the HIFI spectrum toward Orion KL is underway (Crockett et al. 2013b, in preparation), the preliminary results of which are used here to assess the contribution of transitions of other molecules to the observed line profiles.

\subsection{ALMA}

The interferometric observations presented here are part of a Band 6 survey (214--247 GHz) collected by ALMA as part of its science verification.  The full calibrated measurement set is publicly available at https://almascience.nrao.edu/alma-data/science-verification.  The observations were taken on 20 January 2012, with a total of 16 antennas, all 12 m in diameter.  The phase center for the observations was $\alpha_{\textnormal{J2000}} = 05^\textnormal{h} 35^\textnormal{m} 14^\textnormal{s}.35$, $\delta_{\textnormal{J2000}} = -05^\circ 22'35''$.  Callisto was used as the absolute flux calibrator, and the quasar J0607-085 was used as the phase calibrator.  At 226 GHz, the ALMA primary beamwidth is $27.4''$, comparable with Herschel.  The projected baselines ranged from 13 to 202 k$\lambda$.

The two transitions of HDO in the data set were extracted and deconvolved using the Common Astronomy Software Applications (CASA) package\footnote{http://casa.nrao.edu} with the CLEAN algorithm.  Before deconvolution, the continuum as estimated from line-free spectral channels near the HDO transitions was subtracted.  Robust weighting was used with a Briggs parameter of 0.0, and a pixel size of $0.2''$.  After deconvolution, the angular resolution of the image at 225.896 GHz was $1.77'' \times 1.16''$, with a P.A. of -5.6$^\circ$, and for the image at 241.561 GHz, the angular resolution was $1.43'' \times 1.03''$, with a P.A. of -5.5$^\circ$.  The channel width was 488.2 kHz ($\sim$0.65 km s$^{-1}$).  The continuum map used here is available at the ALMA science verification website, and was created using the multi-frequency synthesis CLEAN mode of 30 line-free channels at 230.9 GHz, with a resolution of $1.86'' \times 1.37''$.

\section{Results}

\subsection{Gaussian component fitting}

In this work, we have used a total of 20 transitions of H$_2$$^{18}$O, 14 of H$_2$$^{17}$O, 37 of HDO, 6 of HD$^{18}$O, and 6 of D$_2$O.  These counts exclude any transitions that are judged to be severely blended with transitions from other species, based on the fullband analysis (Crockett et al. 2013b, in preparation) or by inspection of the lineshape.  Energy level diagrams indicating the transitions of H$_2$$^{18}$O, H$_2$$^{17}$O, and HDO used in this study are shown in Figure 1.  Here we do not consider any transitions of H$_2$$^{16}$O, because of the very high optical depth of this species; instead we use the minor isotopologues to infer the total abundance of H$_2$O.

\begin{figure*}
\centering
\includegraphics[width=6.5in]{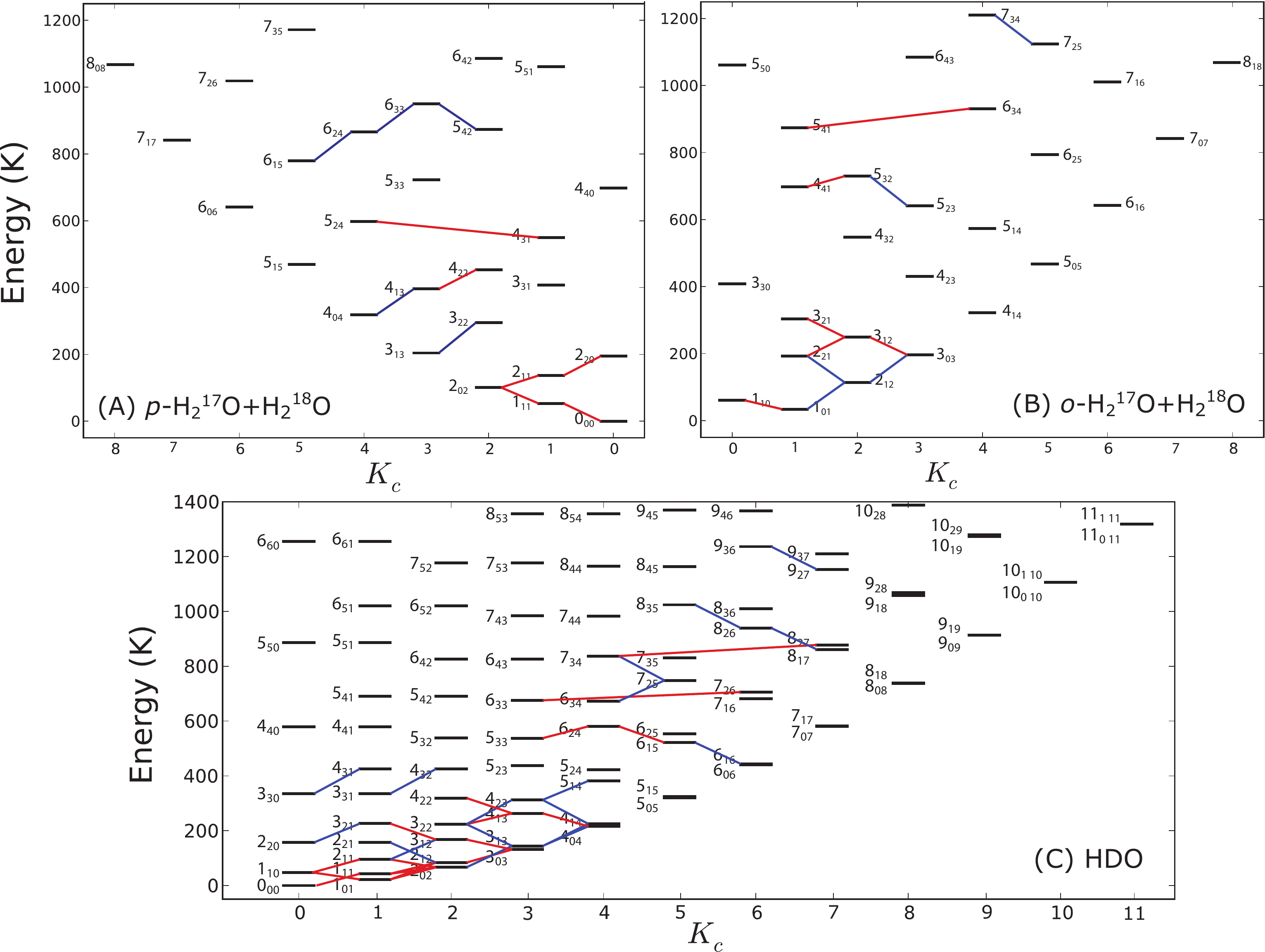}
\caption{Observed transitions of H$_2$$^{18}$O, H$_2$$^{17}$O, and HDO with HIFI.  The lines indicate detected transitions; heavily blended lines (i.e. those judged to be too blended to extract reliable fit Gaussian parameters) are omitted.  Red lines indicate transitions in bands 1-5 of HIFI, while blue lines indicate transitions in bands 6 and 7.  For panels A and B, a transition is indicated if it is clearly detected in either H$_2$$^{18}$O or H$_2$$^{17}$O; for some transitions, due to blends or intensity, both isotopologues are not used in the analysis.}
\end{figure*}

As in previous high-spectral resolution surveys of Orion KL, many molecular transitions exhibit complex lineshapes, corresponding to contributions from multiple spatial components known to exist in this source within the Herschel beam.  Physical and kinematic properties of the three canonical spatial components can be found in Table 1, and are discussed briefly below:

\begin{itemize}
\item Hot Core:  This region has a complex structure, with a number of radio and infrared continuum sources \citep{genzel89, menten95, beuther04}.  It has been proposed that the Hot Core region is heated by the remnants of a recent explosive event \citep{zapata11, bally11, nissen12} rather than active star formation.

\item Compact Ridge:  This is also a structurally complex region, particularly in the observed molecular emission morphologies \citep{friedel08, guelin08, favre11, neill11, brouillet13}, and has also been suggested to have been heated externally \citep{blake87, wang11}.

\item Plateau:  There are two prominent ouflows centered in the Orion KL region \citep{genzel81, genzel89, greenhill98}.  The so-called High-Velocity Flow is oriented in the SE--NW direction and characterized by velocities of up to 150 km s$^{-1}$, while the Low-Velocity Flow ($\Delta v \sim 18$ km s$^{-1}$) is oriented in the NE--SW direction.  In some transitions of water, the blue-shifted wing of the outflow component is found to be in absorption against the strong far-infrared dust continuum \citep{cernicharo06}.
\end{itemize}

\begin{deluxetable}{c c c c c c c}
\tablecaption{Kinematic parameters and physical conditions of the Orion KL spatial components.\tablenotemark{a}}
\tablewidth{0pt}
\tablehead{Component & $\theta_s$ & $v_\textnormal{LSR}$ & $\Delta v$ & $T_\textnormal{kin}$ & $n$(H$_2$) & $N$(H$_2$) \\
 & $('')$ & (km s$^{-1}$) & (km s$^{-1}$) & (K) & (cm$^{-3})$ & (cm$^{-2})$}
\startdata
Hot Core & 5--10 & 3--5 & 5--10 & 150--300 & $10^7$--$10^8$ & $3.1 \times 10^{23}$ \\
Compact Ridge & 5--15 & 7--9 & 3--5 & 80--125 & $10^6$--$10^7$ & $3.9 \times 10^{23}$ \\
Plateau & 20--30 & 6--12 & 20--25 & 100--150 & $10^6$--$10^7$ & $1.8 \times 10^{23}$ \\
\enddata
\tablenotetext{a}{Values compiled from \cite{blake87, tercero10, melnick10, plume12}, and Crockett et al. (2013b, in preparation).}
\end{deluxetable}

\noindent The Orion KL region also has an extended ridge, which consists of cooler ($\sim 60$ K) and less dense ($n$(H$_2$) $\sim 10^5$ cm$^{-3}$) quiescent gas extended across the Herschel beam.  This component has very similar kinematic properties to the Compact Ridge ($v_\textnormal{LSR} = 9$ km s$^{-1}$, $\Delta v = 4$ km s$^{-1}$, \cite{blake87}), and may contribute some flux to the lowest-energy lines, which would most likely be incorporated into the Compact Ridge spectral component.  We expect this contribution to be minor, because H$_2$O transitions are likely very subthermally excited at the physical conditions of the extended ridge.

Each transition was fit with up to four Gaussian components using CLASS, depending on which of the spatial components were detected.  Some low-energy transitions, as in previous measurements, are found to have absorption in the blue-shifted wing; the modeling of these transitions is described in more detail in \S 3.5.  For many of the transitions, particularly the lowest-energy transitions which have contributions from all three spatial components, it was found to be necessary to constrain some of the line positions and widths in order to reduce the number of free parameters.  Where this was needed, the values in Table 1 were used.  For HD$^{18}$O and D$_2$O, the line profiles were well fit by single Gaussian components.  The parameters of the Gaussian fits for all isotopologues can be found in the Appendix (Tables 3-6).

\subsection{Strategy for column density derivations}

Here  the approaches used to derive the H$_2$O and HDO abundances in the different spatial/velocity components within Orion KL are described.  Even for the minor isotopologues analyzed here, many transitions are not optically thin.  For H$_2$$^{18}$O and H$_2$$^{17}$O, the optical depth can be determined by comparing transitions of the two isotopologues, using the following equation:

\begin{equation}
\frac{\Delta T_{\textnormal{mb}}(\textnormal{H}_2\,\!^{18}\textnormal{O})}{\Delta T_{\textnormal{mb}}(\textnormal{H}_2\,\!^{17}\textnormal{O})} =
\frac{J(T_{\textnormal{ex},18})(1-e^{-\tau_{18}})}{J(T_{\textnormal{ex},17})(1-e^{-\tau_{17}})}
= \frac{(1-e^{-\tau_{18}})}{(1-e^{-\tau_{17}})}
\end{equation}

\noindent where we assume the same excitation temperature between transitions with the same quantum numbers of H$_2$$^{18}$O and H$_2$$^{17}$O; therefore, $\tau_{18}/\tau_{17}$ is the H$_2$$^{18}$O/H$_2$$^{17}$O abundance ratio.  We assume $^{16}$O/$^{18}$O = $250 \pm 135$ \citep{tercero10} and $^{18}$O/$^{17}$O = $3.6 \pm 0.7$ \citep{persson07}.  This $^{16}$O/$^{18}$O ratio was derived by \cite{tercero10} from a comparison of $^{16}$OCS and $^{18}$OCS in the Plateau; in the Hot Core and Compact Ridge only lower limits could be estimated for the $^{16}$O/$^{18}$O ratio because of optical depth.  \cite{tercero10} do note that optical depth in the normal isotopologue could still be an issue for the Plateau, so their observations may be consistent with the solar value of 500.  A recent analysis of C$^{18}$O and C$^{17}$O in the Orion KL HIFI spectrum by \cite{plume12} suggested different $^{18}$O/$^{17}$O isotopic ratios between spatial components: they derived a ratio of $3.0_{-1.1}^{+1.2}$ for the Hot Core and $4.1_{-1.3}^{+2.1}$ for the Compact Ridge, within the $1\sigma$ errors of the ratio adopted here ($3.6 \pm 0.7$).  In the Plateau, \cite{plume12} derive a C$^{18}$O/C$^{17}$O ratio of $1.7_{-0.5}^{+0.4}$, which they suggest could be due to isotopically selective photochemistry.  Here, however, we assume the same oxygen isotopic ratios for all spatial components.  

Figure 2 shows the comparison of three corresponding transitions of H$_2$$^{18}$O and H$_2$$^{17}$O.  In the first row, where a transition with $E_\textnormal{up} = 136$ K is shown, it can be seen that much of the flux for low-energy transitions is in the broad Plateau component, but the narrower Hot Core and Compact Ridge components are also clearly visible.  For all three components, a visual inspection reveals that the H$_2$$^{18}$O/H$_2$$^{17}$O flux ratio is significantly less than the assumed abundance ratio of 3.6, indicative of significant optical depth.  For the Hot Core, the flux is \emph{greater} in the H$_2$$^{17}$O transition than in H$_2$$^{18}$O.  This is observed in several lines with $E_\textnormal{up} < 400$ K; this is likely due to foreground extinction of Hot Core emission by the outflow, which is moderately optically thick in H$_2$$^{18}$O; this was previously suggested by \cite{pardo01}.  For the second row in Figure 2, where a transition with $E_\textnormal{up} \sim 450$ K is shown, emission from only the Plateau and Hot Core components is detected.  In the bottom row, where a high-excitation line ($E_\textnormal{up} = 728$ K) is shown, the transitions are well-modeled with a single Gaussian component, attributed to the Hot Core.  In Figure 3, a sample of HDO lines are presented, while in Figure 4 we show the detected lines of HD$^{18}$O and D$_2$O.

\begin{figure}
\begin{center}
\includegraphics[width=6.0in]{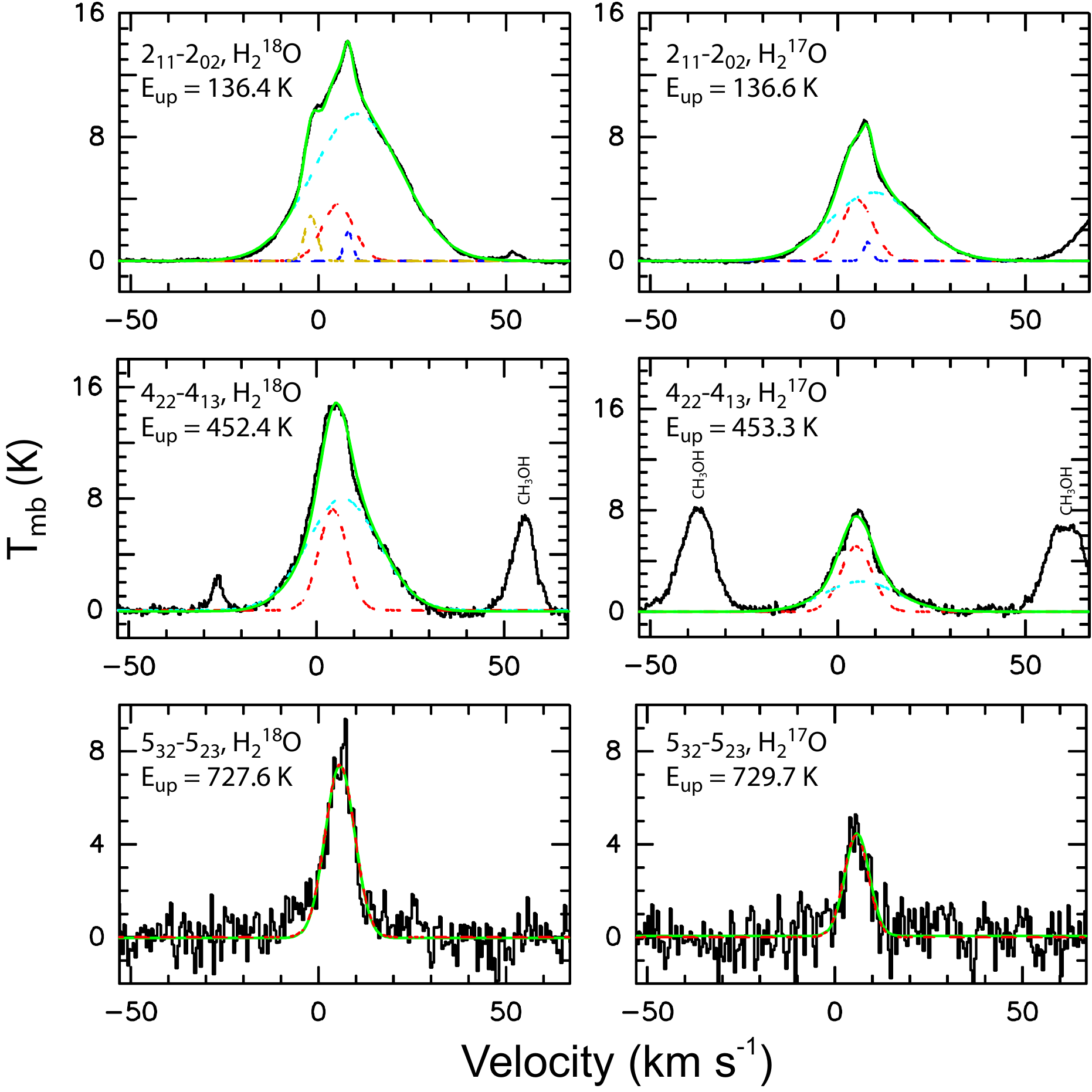}
\caption{Gaussian fits to lines of H$_2$$^{18}$O and H$_2$$^{17}$O.  In each panel, the green curve is the total fit to the data, while the cyan, red, and blue curves indicate the Gaussian components attributed to the outflow, Hot Core, and Compact Ridge, respectively.  In the top-left panel, the yellow curve is the $10_{10,*}-9_{9,*}$ multiplet of CH$_3$OCH$_3$ from the HIFI fullband model.  The spectra (in black) in this figure are continuum-subtracted.}
\end{center}
\end{figure}

\begin{figure}
\begin{center}
\includegraphics[width=6.0in]{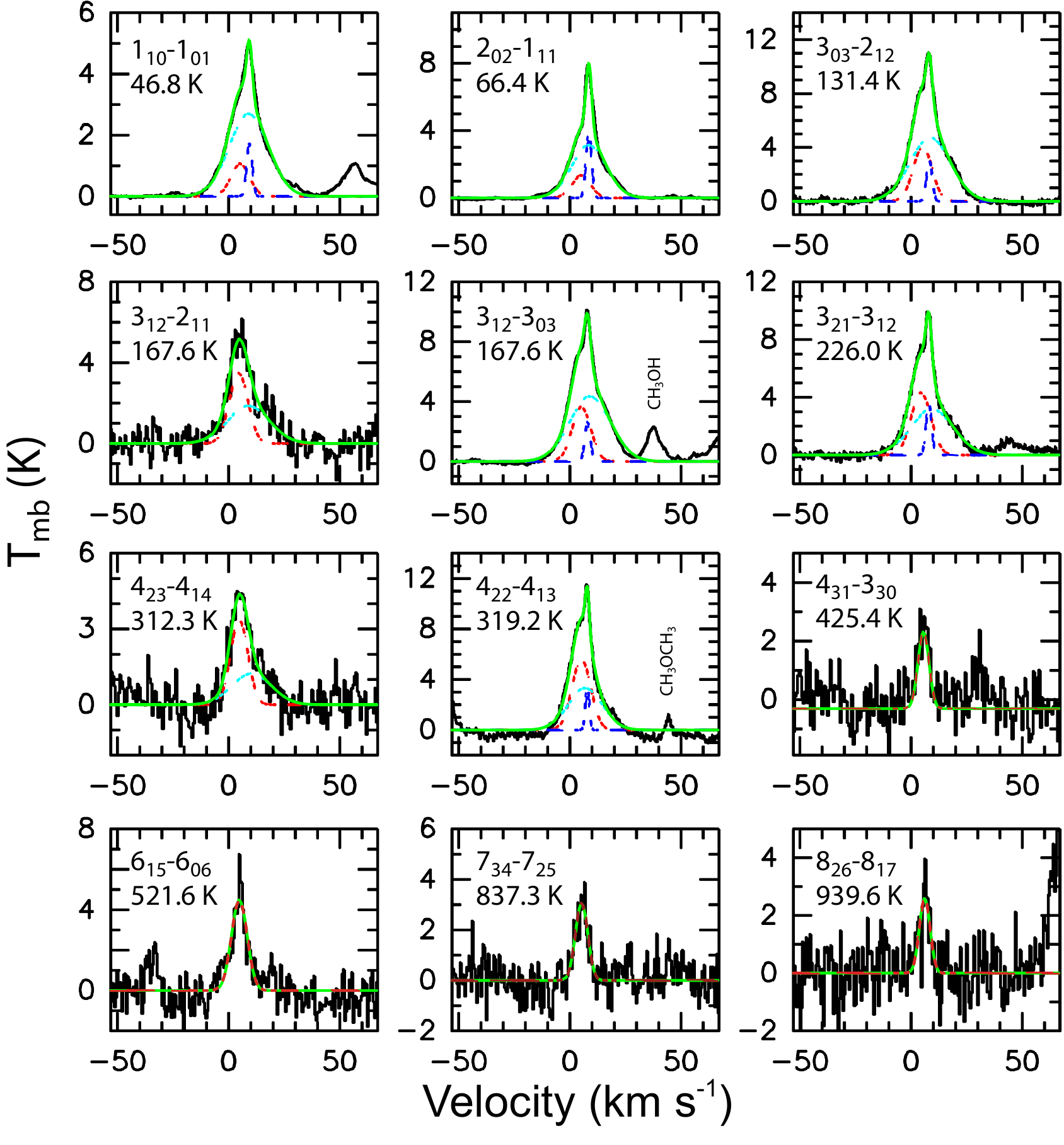}
\caption{Gaussian fits to a selection of HDO transitions detected by HIFI.  In each panel, the green curve is the total fit to the data, while the cyan, red, and blue curves indicate the Gaussian components attributed to the outflow, Hot Core, and Compact Ridge spatial components, respectively.  The quantum numbers and the upper-state energy of each transition are specified.  The spectra (in black) in this figure are continuum-subtracted.}
\end{center}
\end{figure}

\begin{figure}
\begin{center}
\includegraphics[width=5.0in]{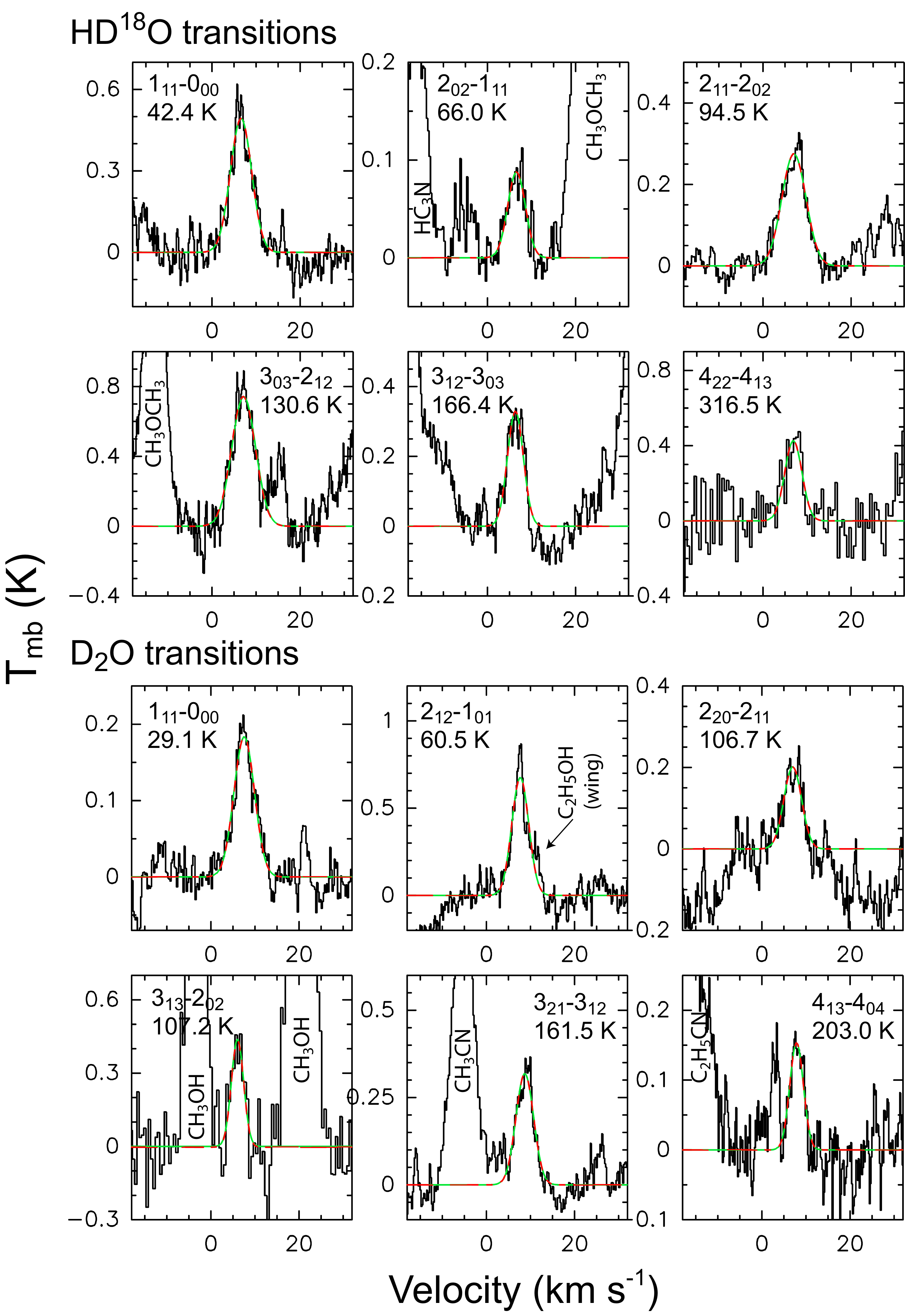}
\caption{Detected transitions of HD$^{18}$O and D$_2$O.  The red curves indicate single-Gaussian fits to the data.  The quantum numbers and upper-state energy are indicated for each transition.  The spectra (in black) in this figure are continuum-subtracted.}
\end{center}
\end{figure}

Because the H$_2$ density within Orion KL is likely lower than is required to collisionally thermalize all of the observed transitions, the level populations are expected to deviate significantly from local thermodynamic equilibrium (LTE).  Additionally, the excitation of water is strongly influenced by the local far-infrared radiation field \citep{jacq90, cernicharo06, melnick10, vandishoeck11}.  We have therefore included a background continuum field based on far-infrared observations of Orion KL, which is presented in Figure 5.  Further information on this continuum can be found in Crockett et al. (2013a, in preparation).  The observations derive from the continuum level measured by HIFI in the Orion KL fullband survey for $\lambda =$ 600--160 $\upmu$m, and from Infrared Space Observatory surveys for shorter wavelengths \citep{vandishoeck98, lerate06}.  The ISO observations are scaled to match those from HIFI at their intersection wavelength ($160$ $\upmu$m).  Because the HIFI beam at this wavelength ($11''$) is smaller than that of ISO-LWS ($\sim$80$''$), the higher continuum flux measured by HIFI is attributed to greater beam dilution in ISO.  The resulting continuum (in black in Figure 5) is referred to here as the ``observed continuum.''  

A recent study of the excitation of H$_2$S in the Orion Hot Core with the HIFI fullband survey (Crockett et al. 2013a, in prepration) has found that reproducing the observed line fluxes, particularly for the highest energy levels, requires an enhancement of a factor of 8 for $\lambda < 100$ $\upmu$m above the observed continuum in Figure 5, a possible indication of hidden luminosity from hot dust in the Hot Core not directly detectable due to high optical depth but important in the excitation of hydride molecules with transitions in the far-infrared.  The Hot Core has been previously suggested to have high optical depth in the far-IR on the basis of modeling of high-excitation NH$_3$ \citep{hermsen88} and HDO \citep{jacq90} transitions.  As will be discussed further in \S 3.3 below, better agreement is found with the observed line fluxes of H$_2$O and HDO in the Hot Core when the continuum is enhanced by a factor of 3 in the far-IR.  This continuum is plotted in green in Figure 5 and referred to as the ``enhanced continuum'' in this work.  For $\lambda > 100$ $\upmu$m, the dust optical depth is lower, so it is less likely that the true continuum field seen by the molecular gas is significantly enhanced over the observed continuum.  The H$_2$O and HDO excitation is less sensitive to radiative pumping at longer wavelengths, so this makes little impact on the derived abundances.  For the Compact Ridge and Plateau spatial components, far-infrared excitation is also important, and for these components the observed continuum in Figure 5 is used.

\begin{figure}
\centering
\includegraphics[width=6.0in]{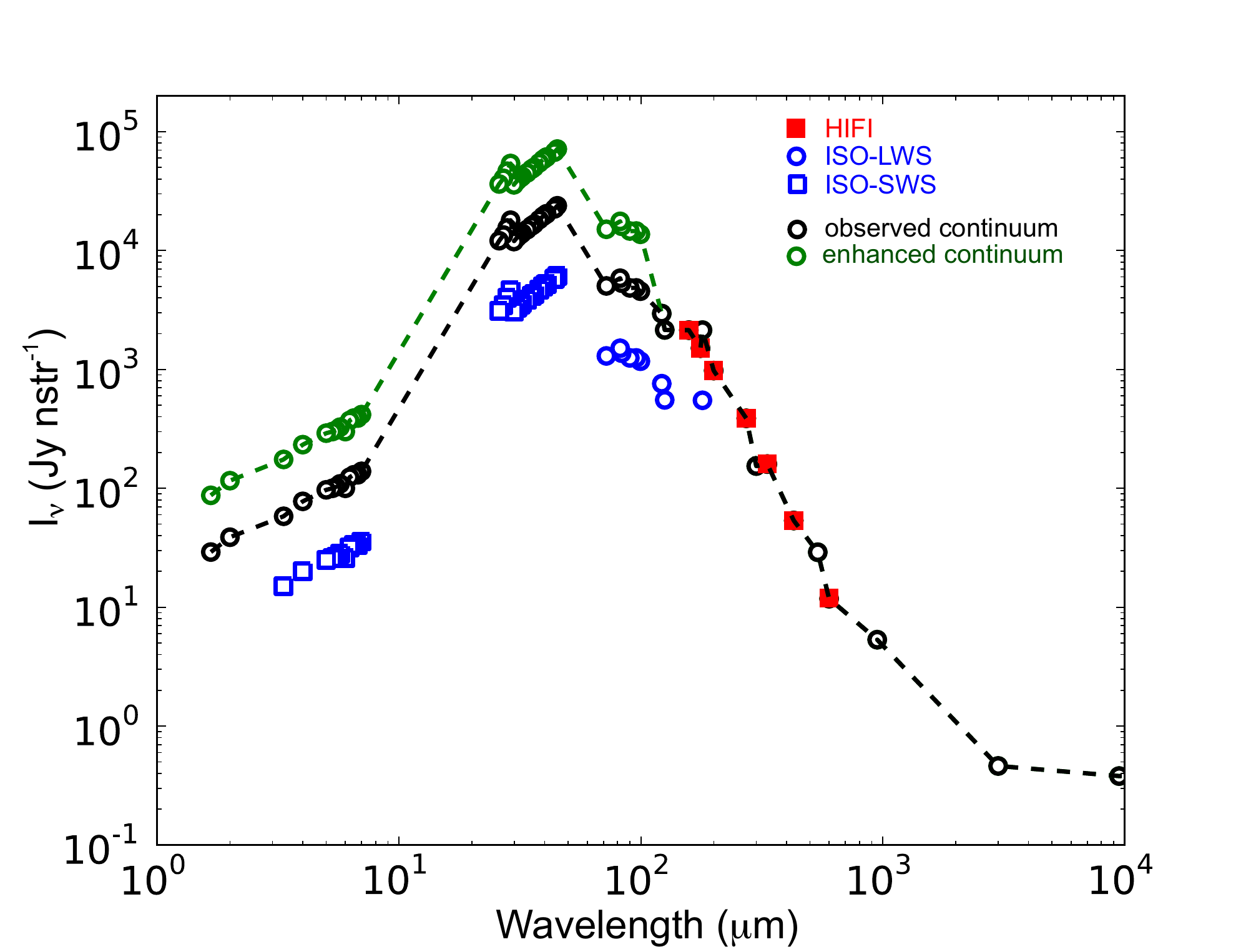}
\caption{Orion KL continuum radiation field used for modeling of H$_2$O and HDO emission.  The blue and red points indicate measurements, as indicated.  The ISO-LWS measurements are from \cite{lerate06}; ISO-SWS measurements are from \cite{vandishoeck98}; and the HIFI measurements are from the Orion KL fullband survey \citep{bergin10}.  The black and green curves show the ``observed'' and ``enhanced'' continuum fields used in RADEX modeling.  See the text and Crockett et al. (2013a, in preparation) for further information.}
\end{figure}

We have used two different approaches to derive the column densities of H$_2$O and HDO; the method used for a given spatial component depends on the reliability of the optical depth estimates for each component and the number of transitions observed to emit from the component.  The first approach is to directly sum the populations of each observed level \citep{goldsmith97, plume12}.  Figure 1 shows that particularly for low-lying ($E < 400$ K) energy levels where most of the population is found, transitions are detected originating from most levels.  From each transition, the population in the upper state can be derived using \citep{goldsmith99}

\begin{equation}
N_u = \frac{1.67 \times 10^{17} W g_u}{\nu S_{\textnormal{ij}} \mu^2 \eta_{\textnormal{bf}}} \frac{\tau_l}{1-\textnormal{e}^{-\tau_l}}
\end{equation}

\noindent where $N_u$ is the upper state column density in cm$^{-2}$, $W$ is the integrated flux in K km s$^{-1}$, $g_u$ is the upper state degeneracy, $\nu$ is the frequency in MHz, $S_{\textnormal{ij}}\mu^2$ the line strength in debye$^2$, $\eta_\textnormal{bf}$ the beam dilution factor, and $\tau_l$ the line optical depth.  The column densities in individual levels derived by equation (2) are independent of the excitation mechanism, whether through collisions or radiative excitation.  In order to derive a total column density, the following equation is used:

\begin{equation}
N_\textnormal{total} = f_c \sum N_\textnormal{observed}
\end{equation}

\noindent In this equation, $f_c$ is a correction factor to account for the population that is located in levels that cannot be derived by equation (2).  These factors are calculated from 1-D large velocity gradient calculations using the publicly available RADEX code \citep{vandertak07}.  The physical parameters from Table 1 are used for these calculations.  Where this approach, referred to here as the population summation method, is not possible, we have used the RADEX code to derive the column density and physical parameters that best reproduce the observed measurements, which will be described in more detail in the sections to follow.

These models use collisional rates for isotopologues of water with H$_2$ from the LAMDA database \citep{schoier05}.  For H$_2$$^{18}$O and H$_2$$^{17}$O, the rates from \cite{daniel11} for collisions of H$_2$O with H$_2$ are used, while for HD$^{16}$O and HD$^{18}$O we use the rates of \cite{faure12} for HDO.  For all isotopologues, the rates were calculated for collisions with both \emph{o}-H$_2$ and \emph{p}-H$_2$, and a thermal \emph{ortho}/\emph{para} H$_2$ ratio is assumed in all models.  While transitions of HDO with energies up to 1200 K are detected, the available collision rates for HDO only include energy levels up to $E = 450$ K.  At the present time, therefore, we cannot model the highest-energy transitions of HDO detected by HIFI.  Additionally, the models presented here do not include the effect of radiative pumping through vibration-rotation transitions.  If this excitation pathway is important, it could change the physical parameters derived in this study.  However, for each component enough transitions are detected that the abundance is well constrained, despite uncertainty in the precise excitation mechanism.

In the following subsections, we discuss in detail the derivation of the H$_2$O and HDO column densities in each component, which are summarized in Table 2.  In this table we present the column density of HDO relative to both H$_2$$^{18}$O and H$_2$$^{16}$O.  For most components, the HDO/H$_2$O ratio is dependent on the $^{16}$O/$^{18}$O ratio.  The uncertainty in this ratio is therefore a major contributor to the uncertainty in the absolute HDO/H$_2$O ratio for each component, so for comparison of the deuterium fractionation between the different components within Orion KL, the [HD$^{16}$O]/[H$_2$$^{18}$O] ratio is more representative of the relative uncertainties.

\begin{deluxetable}{c c c c c c c}
\tabletypesize{\footnotesize}
\tablewidth{0pt}
\tablecaption{Summary of H$_2$O and HDO column densities in Orion KL.}
\tablehead{Component & $\theta_s$ & $N$(H$_2$$^{18}$O) & $\chi$(H$_2$O)\tablenotemark{a} & $N$(HDO) & [HD$^{16}$O]/[H$_2$$^{18}$O] & [HDO]/[H$_2$O] \\
& $('')$ & (cm$^{-2}$) & & (cm$^{-2}$) }
\startdata
Hot Core (7 km s$^{-1}$) & 2 & $8.0_{-4.0}^{+8.0} \times 10^{17}$ & $6.5_{-4.8}^{+7.3} \times 10^{-4}$ & $6.0_{-3.6}^{+3.6} \times 10^{17}$ & $0.75_{-0.58}^{+0.88}$ & $3.0_{-1.7}^{+3.1} \times 10^{-3}$ \\[1ex]

Hot Core (5 km s$^{-1}$) & 5 & $\ge 9.0 \times 10^{15}$ & $\ge 7.3 \times 10^{-6}$ & $\ge 6.2 \times 10^{15}$ & 0.69 & $2.8 \times 10^{-3}$ \\[1ex]

Compact Ridge & 6 & $4.1_{-0.9}^{+1.0} \times 10^{15}$ & $2.6_{-1.5}^{+1.6} \times 10^{-6}$ & $3.9_{-1.2}^{+2.9} \times 10^{15}$ & $0.95_{-0.36}^{+0.74}$ & $3.8_{-2.5}^{+3.6} \times 10^{-3}$ \\[1ex]

Plateau (emission) & 30 & $3.5_{-0.6}^{+0.6} \times 10^{15}$ & $4.8_{-2.8}^{+2.8} \times 10^{-6}$ & $1.23_{-0.25}^{+0.25} \times 10^{15}$ & $0.35_{-0.09}^{+0.09}$ & $1.4_{-0.8}^{+0.8} \times 10^{-3}$ \\[1ex]

Absorbing gas & $\theta_b$\tablenotemark{b} & $9.8_{-2.9}^{+4.0} \times 10^{14}$ & $2.7_{-1.6}^{+1.6} \times 10^{-6}$ & $5.5_{-0.7}^{+0.6} \times 10^{13}$ & $0.056_{-0.013}^{+0.014}$ & $2.2_{-1.3}^{+1.3} \times 10^{-4}$

\enddata
\tablenotetext{a}{Assuming H$_2$ column densities from Table 1, except for the absorption component, where $N$(H$_2$) = $9.0 \times 10^{22}$ cm$^{-2}$ \citep{phillips10} is assumed.}
\tablenotetext{b}{We assume that the absorbing gas fills the Herschel beam at each frequency.}
\end{deluxetable}

\subsection{Hot Core}

\subsubsection{HDO and HD$^{18}$O}

The detection of HD$^{18}$O toward Orion KL was first reported by \cite{bergin10}, to date the only time this species has been identified in the interstellar medium.  This detection implies a region with a very high HDO column density  within the Herschel beam.  HDO emission from the component that is emissive in HD$^{18}$O must have very high optical depth in the corresponding HD$^{16}$O transitions (comparing Figures 3 and 4).  The HD$^{18}$O lines have an average $v_\textnormal{LSR} = 6.7$ km s$^{-1}$ and $\Delta v = 5.4$ km s$^{-1}$, both of which are intermediate between the canonical parameters for the Compact Ridge and Hot Core given in Table 1.  This detection was initially reported as tenative by \cite{bergin10}.  The fullband model to the HIFI spectrum (Crockett et al. 2013b, in preparation) does not have any substantial ($> 0.1$ K) transitions of other species coincident in frequency with any of these six transitions.  An LTE model to the six transitions of HD$^{18}$O does not predict any lines to be emissive that are missing; due to the high line density of the Orion KL spectrum, there are several potentially emissive transitions that lie under strong lines of other species.  We conclude that the assignment of these six transitions to HD$^{18}$O is correct.  The high HDO column density implied by this detection requires one of two explanations:  either the HD$^{18}$O-emitting component has a high [HDO]/[H$_2$O] ratio as compared to the ``normal'' D/H ratios previously found for water and other species in Orion KL ($\sim 10^{-3}$--$10^{-2}$), or this component also has a high H$_2$O abundance.  The $2_{20}-2_{21}$ transition of HDO at 10.3 GHz was detected by \cite{petuchowski88} with similar kinetic parameters as the HD$^{18}$O transitions.  This detection was surprising considering the low Einstein \emph{A} coefficent of this transition ($3.6 \times 10^{-9}$ s$^{-1}$) and was interpreted as evidence of a high HDO abundance in a highly excited clump of gas.

In order to determine the spatial origin of the HD$^{18}$O emission, we use two transitions of HDO that were detected in the Orion KL ALMA survey: the $3_{12}-2_{21}$ transition at 225896.7 MHz ($E_\textnormal{up} = 167.6$ K, $S_\textnormal{ij} \mu^2 = 0.69$ D$^2$) and the $2_{11}-2_{12}$ transition at 241561.6 MHz ($E_\textnormal{up} = 95.2$ K, $S_\textnormal{ij} \mu^2 = 0.36$ D$^2$).  There is also a third potentially detectable transition of HDO in the dataset (the $7_{34}-6_{43}$ transition at 241.973 GHz), but this line is  blended with a strong transition of C$_2$H$_5$CN.  Figure 6 shows images of these two transitions, integrated over $\sim 2.5$ km s$^{-1}$ velocity widths.  Both of these transitions appear to be free of significant emission from other molecules, and the emission morphologies of the two transitions are very similar.  The 225 GHz transition was found in LVG modeling by \cite{faure12} to exhibit a moderate population inversion ($|\tau| \le 1$) under high densities and temperatures like the conditions within Orion KL, but the 241 GHz transition did not.  In the first row, in the velocity range of 2.8--5.7 km s$^{-1}$, the strongest emission comes from the Hot Core region, near the region of strongest 230 GHz continuum emission, with a second component near the IRc7 infrared continuum source.  In the second row, it can be seen that the strongest emission in the 6.3--8.9 km s$^{-1}$ velocity range is located about $1''$ south of the dust continuum peak, centered at $\alpha_\textnormal{J2000}$ = 05$^\textnormal{h}$35$^\textnormal{m}$14$^\textnormal{s}$.54, $\delta_\textnormal{J2000}$ = -05$^\circ 22'33''$.  This spatial component peaks at a velocity of 7 km s$^{-1}$, in agreement with the velocity of the HD$^{18}$O lines in HIFI.  Lastly, in the third row, showing velocities between 9.5--11.5 km s$^{-1}$, in addition to the Hot Core emission (which is the red wing of the 7 km s$^{-1}$ component), emission from the Compact Ridge (to the southwest of the Hot Core) and a clump to the northwest can be seen.  The emission in this velocity range is weaker than in the other two rows; note that the color scale is more sensitive by a factor of $\sim 3$ in the third row.  The ALMA observation was performed without zero-spacing information.  Therefore, to judge the degree to which these observations may be missing extended emission, we compared the line fluxes to the single-dish observations of these two transitions with the IRAM 30 m telescope by \cite{jacq90}.  The ALMA images were smoothed to a spatial resolution of $10.5''$, the beamwidth at the 30 m telescope at this frequency, and found that 80--90\% of the flux of these two transitions is recovered by ALMA.

\begin{figure}
\begin{center}
\includegraphics[width=5.5in]{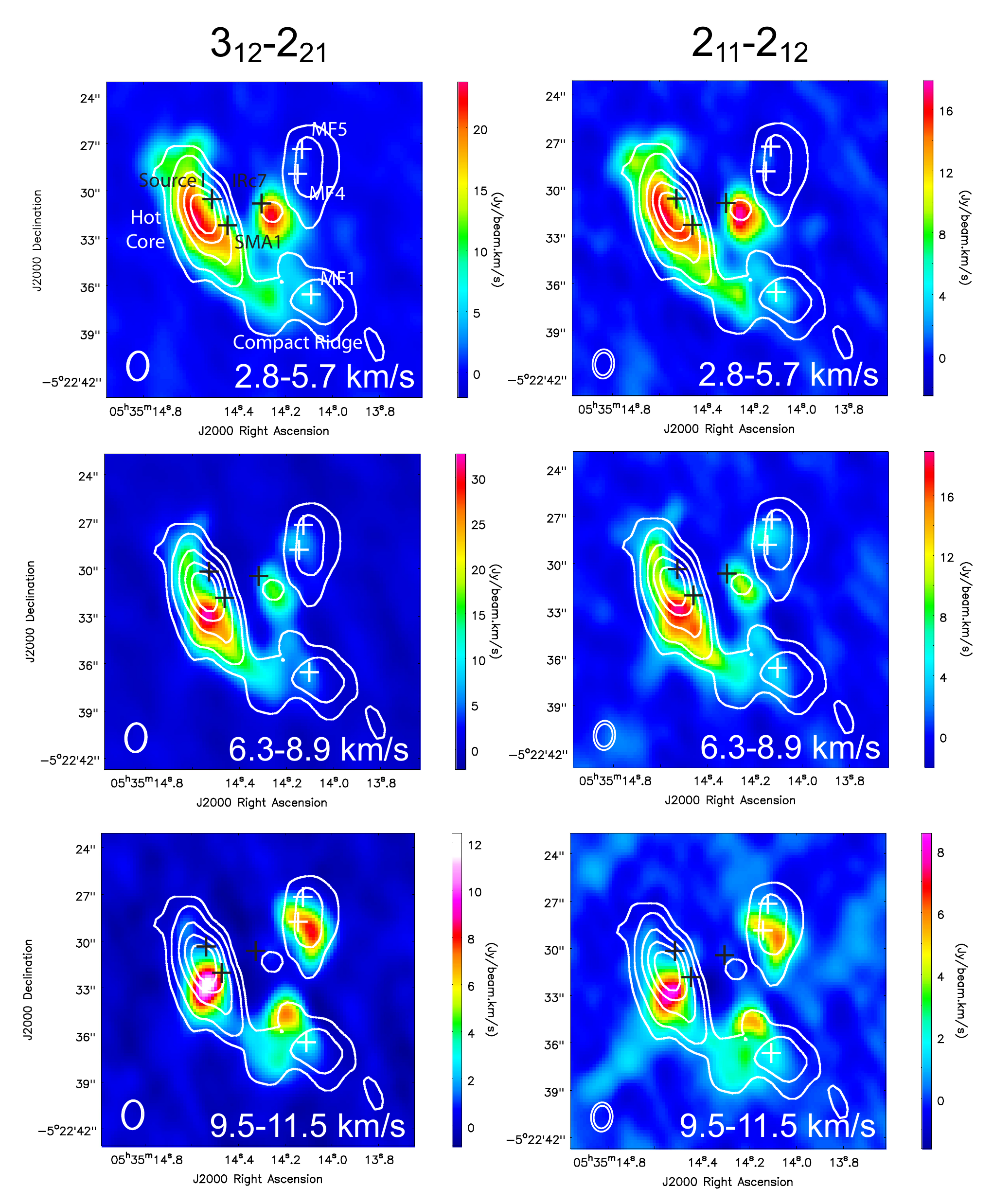}
\caption{Emission of two transitions of HDO toward Orion KL as measured by ALMA.  In each panel, the integrated HDO emission over the given velocity range is in color scale, while the white contours (levels (0.1, 0.2, 0.4, 0.6, 0.8) $\times$ 1.34 Jy beam$^{-1}$) indicate the continuum emission level.  The black crosses indicate three of the continuum sources in this region: (from left to right) source I, SMA1, and IRc7.  The white crosses indicate peaks in methyl formate emission in the nomenclature of \cite{favre11}.  The ovals in the lower left corner of each panel indicate the synthesized beam.}
\end{center}
\end{figure}

The agreement between the velocity of the region with strongest HDO emission in Orion KL in the ALMA images in Figure 6 and the velocity of the detected HD$^{18}$O lines by HIFI allows us to assign the HD$^{18}$O emission to the clump to the south of the Hot Core region.  Motivated by these maps, we model the water emission in the Hot Core with two spatial/velocity components, one for the 7 km s$^{-1}$ clump and a second centered at the canonical Hot Core velocity of 5 km s$^{-1}$, consisting of the emission components in the top row of Figure 6.  We assume that all of the HD$^{18}$O emission comes from the 7 km s$^{-1}$ component, and we begin with models to derive the HDO abundance in this component.  In Figure 7, two single-component models of the Hot Core emission of HDO and HD$^{18}$O are shown.  The points indicate the fluxes for each transition; for HD$^{16}$O, because the Hot Core is fit as a single Gaussian component, it represents the total flux summed over the two components.  These models are calculated with a kinetic temperature of 200 K, an H$_2$ density of 10$^8$ cm$^{-3}$, and the enhanced background continuum field shown in Figure 5.  If the observed continuum is used instead, an equally good fit can be obtained with a higher H$_2$ density ($10^9$ cm$^{-3}$), which may be reasonable over a small region.  As discussed in \S 3.3.2 below, the high-energy ($E_u > 500$ K) H$_2$O lines are best modeled with the enhanced continuum, so for consistency, we also use the enhanced field for the models in Figure 7.  The derived HDO column density is insensitive to these two excitation scenarios.

\begin{figure}
\begin{center}
\resizebox{\hsize}{!}{\includegraphics{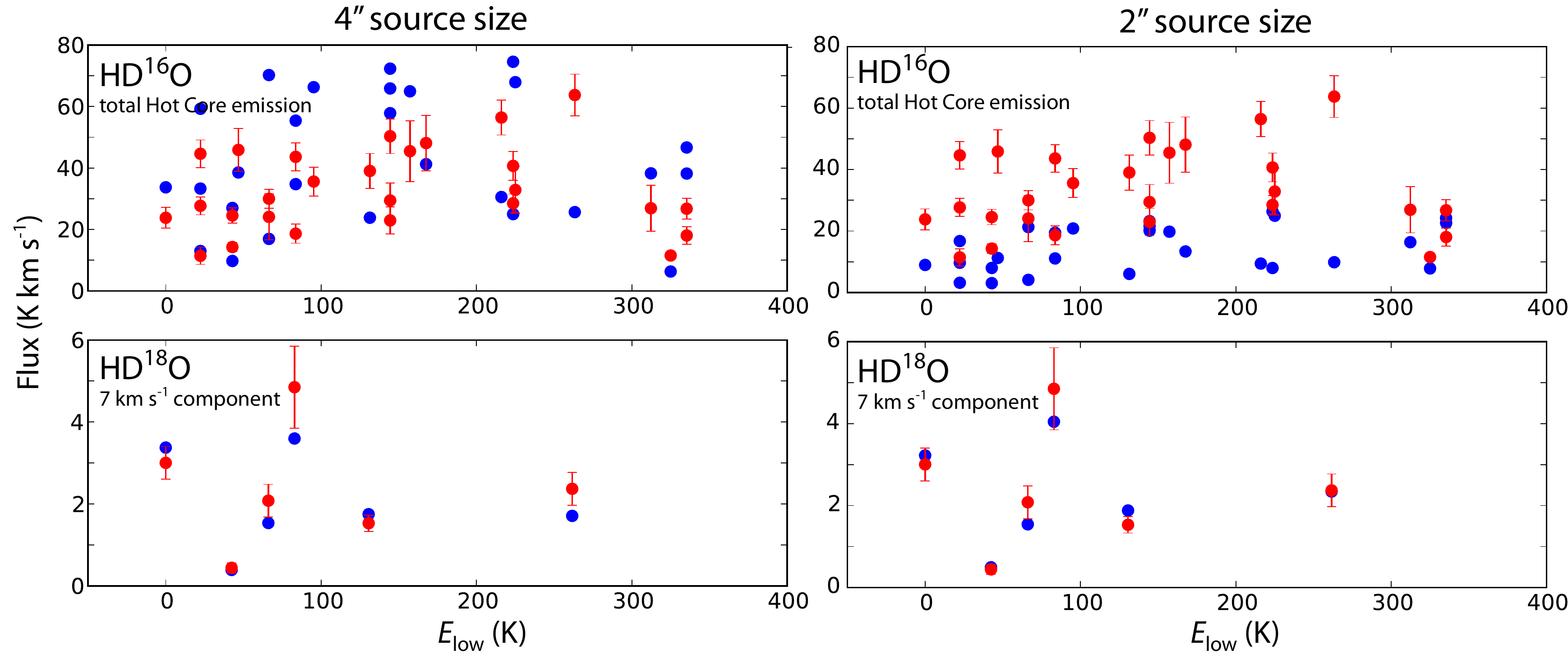}}
\caption{RADEX models of HDO and HD$^{18}$O to the 7 km s$^{-1}$ component in the Hot Core.  In all panels, the red circles (with error bars) represent the observed line fluxes as a function of lower-state energy, while the blue circles are the integrated line fluxes calculated by the models.  The left column is with a source size of $4''$ and $N$(HD$^{18}$O) = $4.0 \times 10^{14}$ cm$^{-2}$, while the model in the right column has a source size of $2''$ and $N$(HD$^{18}$O) = $2.4 \times 10^{15}$ cm$^{-2}$. An HD$^{16}$O/HD$^{18}$O ratio of 250 is assumed.  Both models use $T_{\textnormal{kin}}$ = 200 K, $n$(H$_2$) = $10^8$ cm$^{-3}$, and the enhanced far-infrared radiation field as described in the text.}
\end{center}
\end{figure}

The two models in Figure 7 differ in the size of the emitting region.  For a source size of $4''$ (left column), the fluxes of the HD$^{18}$O lines are well reproduced, but many of the lines of HD$^{16}$O are overpredicted, some by as much as a factor of 3.  While the outflow component of the HDO transitions could have significant optical depth in some lines and therefore could hide the Hot Core component in some of these lines \citep{pardo01}, it is unlikely that the extinction is this significant in all of these transitions, particularly as the Plateau component is weak in the higher-energy lines.  Therefore, the most likely explanation is that the emitting region responsible for the HD$^{18}$O lines is smaller than $4''$, and the HDO emission from the same region is more optically thick.  The right panel shows that with a source size of $2''$ and a column density $N$(HD$^{18}$O) = $2.4 \times 10^{15}$ cm$^{-2}$, and the same excitation parameters otherwise, the HD$^{18}$O lines are still well modeled, but the optical depth in the HD$^{16}$O transitions is high enough to keep the these lines from being overpredicted.  A source size of $2''$ also agrees well with the size of the bright 7 km s$^{-1}$ clump in the HDO ALMA images in Figure 6.  Therefore, we adopt $2''$ as the size of the HD$^{18}$O emitting region, and $(2.4 \pm 0.6) \times 10^{15}$ cm$^{-2}$ as the HD$^{18}$O column density.  This 25\% uncertainty is attributed primarily to the uncertainty in the excitation of the HD$^{18}$O transitions.  This leads to an HDO column density of $(6.0 \pm 3.6) \times 10^{17}$ cm$^{-2}$.

In the HD$^{16}$O panel of the $2''$ model in Figure 7, the difference between the observed fluxes and the calculated fluxes of the 7 km s$^{-1}$ component is attributed to the second Hot Core component (centered at 5 km s$^{-1}$).  We adopt a source size of $5''$ for this component, and use the population summation method described in \S 3.2.  We subtract from the observed flux of each line the flux predicted by the $2''$ model to the 7 km s$^{-1}$ component in Figure 7, and then apply equation (1) to derive an upper state column density for each transition, assuming that the emission from this component is optically thin.  There are several cases where two transitions with the same upper state are observed; these line pairs suggest that some transitions in this component have moderate optical depth.  For these cases, we use the transition with lower optical depth to estimate the population in that level. However, as we do not have information about the optical depth for most levels of HDO, this column density should be viewed as a lower limit.  We calculate a correction factor of 1.42, which is derived as described above using RADEX, assuming $T_\textnormal{kin} = 200$ K, $n$(H$_2$) = $10^8$ cm$^{-2}$, and using the enhanced continuum field.  With this, an HDO column density of $6.2 \times 10^{15}$ cm$^{-2}$ for the 5 km s$^{-1}$ Hot Core component is derived.

\subsubsection{H$_2$$^{18}$O and H$_2$$^{17}$O}

The analysis of H$_2$O in the Hot Core is complicated by the fact that a comparison of H$_2$$^{18}$O and H$_2$$^{17}$O, as described above, shows that many of the lower-energy transitions are very optically thick, so they contain little to no information on the column density in those levels; also, as noted above, the Plateau could be attenuating the Hot Core emission in some lines.  The only transitions of H$_2$$^{18}$O or H$_2$$^{17}$O that might be optically thin are the high-energy lines, so we first turn our attention to the highest-energy transitions before returning to discuss the lower-energy, optically thick lines.  The high-energy transitions (here defined as $E_\textnormal{up} > 500$ K) that fall in the HIFI bandwidth can be broadly segregated into two sets based on their line strengths:  $\Delta J$ = 0 transitions, which have high line strengths ($S_{\textnormal{ij}} \mu^2 > 10$ D$^2$), and $\Delta J$ = 1 transitions, which are are considerably weaker ($S_{\textnormal{ij}} \mu^2 < 3$ D$^2$).  These transitions therefore span a wide range in optical depth.  As with the HD$^{18}$O transitions, the detection of these low-$S_\textnormal{ij}\mu^2$, high-energy H$_2$$^{18}$O transitions indicate a component with a high H$_2$O abundance.

Figure 8 shows transitions of H$_2$$^{18}$O and H$_2$$^{17}$O with $E_\textnormal{up} > 500$ K.  Several of the lines, particularly the lines of H$_2$$^{17}$O, are only marginally detected, and three are blended with lines of CH$_3$OH.  The two overlaid models have source sizes of $4''$ (dashed red lines) and $2''$ (solid red lines).  Both models have $T_\textnormal{kin} = 200$ K and $n$(H$_2$) = $10^8$ cm$^{-3}$, and the enhanced continuum field from Figure 5.  For both models, the low-$S_\textnormal{ij}\mu^2$ transitions are well reproduced, with the exception of the $6_{33}-5_{42}$ transition; this line is not well reproduced with any model that does not overpredict other transitions substantially, so this line may be blended with an unidentified transition from another molecule.  However, for the $4''$ source size model, with the column density required to reproduce the flux of the low-$S_\textnormal{ij}\mu^2$ transitions, the high-$S_\textnormal{ij}\mu^2$ transitions are overpredicted.  The $2''$ model, alternatively, is in better agreement.  This model has a H$_2$$^{18}$O column density of $8.0 \times 10^{17}$ cm$^{-2}$, which implies a H$_2$$^{16}$O column density of $2.0 \times 10^{20}$ cm$^{-2}$.  We assume a factor of 2 uncertainty in the H$_2$$^{18}$O column density:  in the models presented in Figure 8, $\sim$90\% of the population is located in states with $E_\textnormal{up} <$ 500 K, but, due to high optical depth, there is little to no direct sensitivity to the population in these levels.  Using an H$_2$ column density of $3.1 \times 10^{23}$ cm$^{-2}$ \citep{plume12}, this corresponds to an H$_2$O abundance relative to H$_2$ of $6.5_{-4.8}^{+7.3} \times 10^{-4}$, making H$_2$O the predominant form of oxygen: the H$_2$O column density we derive relative to H ($\approx 2N$(H$_2$)) is $3.25 \times 10^{-4}$, while the Orion Nebula has been found to have total [O]/[H] $\sim 4 \times 10^{-4}$ \citep{wilson94, rubin91, baldwin91}.  It should be noted, however, that the value used for the H$_2$ column density was derived for the Hot Core as a whole, and may be higher in the localized $2''$ region under consideration.  Interferometric studies deriving H$_2$ column densities from millimeter dust emission have found $N$(H$_2$) $\ge 10^{24}$ cm$^{-2}$ over small spatial scales in the center of the Hot Core region \citep{blake96, beuther04, favre11}.

\begin{figure}
\begin{center}
\includegraphics[width=4.5in]{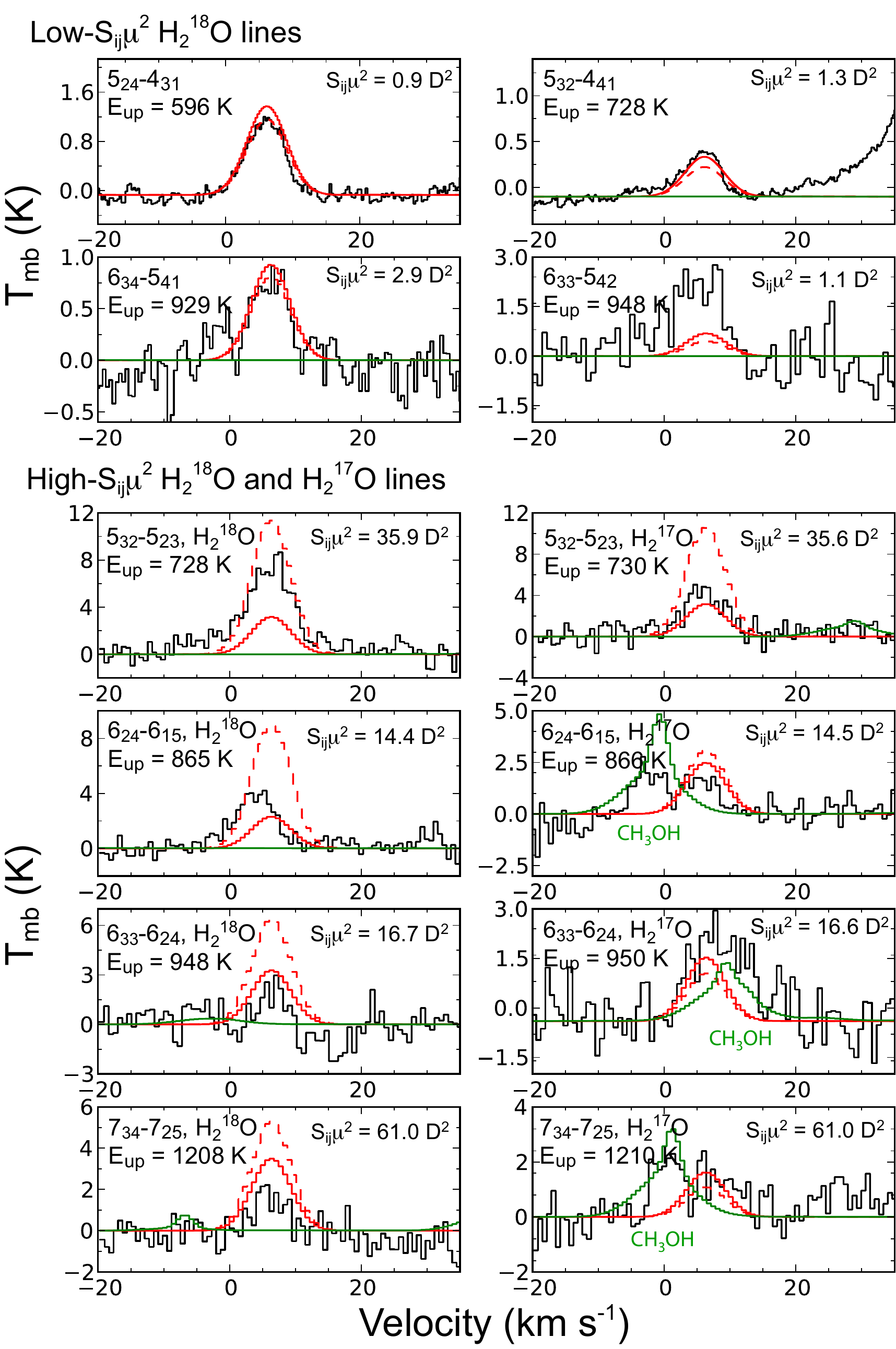}
\caption{High-excitation ($E_{\textnormal{up}} > 500$ K) transitions of H$_2$$^{18}$O and H$_2$$^{17}$O detected (or tentatively detected) toward the Hot Core.  In each panel, the black curves are the data (smoothed to $\sim 1.5$ km s$^{-1}$ to reduce the noise level); the dashed red lines are a RADEX model with $\theta_s = 4''$ and $N$(H$_2$$^{18}$O) = $2.0 \times 10^{17}$ cm$^{-2}$; and the solid red lines are a RADEX model with $\theta_s = 2''$ and $N$(H$_2$$^{18}$O) = $8.0 \times 10^{17}$ cm$^{-2}$; and the green lines are the HIFI fullband model (Crockett et al. 2013b, in preparation).  The assumed kinematic parameters for the synthetic line profiles are $v_\textnormal{LSR} = 6.0$ km s$^{-1}$ and $\Delta v = 7.0$ km s$^{-1}$. Both models have $T_\textnormal{kin} = 200$ K, $n$(H$_2$) = $10^8$ cm$^{-3}$, and an enhanced far-IR radiation field as described in the text.}
\end{center}
\end{figure}

In the models in Figure 8, we assume a \emph{ortho}:\emph{para} ratio of 3 for the H$_2$O isotopologues.  We examined the effect of the \emph{ortho}:\emph{para} ratio on our H$_2$O models by instead assuming a ratio of 1.5 and re-running the models in Figure 8:  we find that the fit is slightly worse (particularly on the low-$S_\textnormal{ij} \mu^2$ lines), but only marginally, so these models are formally consistent with either \emph{ortho}:\emph{para} ratio.  The adopted \emph{ortho}:\emph{para} ratio does not make a dramatic difference in the total H$_2$O column density.  We also assume that the highly excited $2''$ H$_2$O component is cospatial with the $2''$ HDO component from which HD$^{18}$O emission is detected.  In the models shown in Figure 8, kinematic parameters of $v_\textnormal{LSR} = 6.0$ km s$^{-1}$ and $\Delta v = 7.0$ km s$^{-1}$ are assumed for the plotted Gaussians, close to the parameters of the HD$^{18}$O transitions.  It can be seen that the line profiles are generally in good agreement with the observations.  The analyses of HDO and H$_2$O above indicate that both molecules arise in a small clump with high abundance in the Orion Hot Core region, as evidenced by the detection of weak transitions (the rare HD$^{18}$O isotopologue, and the low-$S_\textnormal{ij} \mu^2$ transitions of H$_2$$^{18}$O).  Therefore, the most likely explanation is that they are cospatial.

Just as for HDO, this $2''$ component does not explain all of the flux for the lower-energy H$_2$$^{18}$O and H$_2$$^{17}$O transitions.  As in \S 3.3.1, we assume that the remainder of the flux is attributed to the rest of the Hot Core region (the 5 km s$^{-1}$ component), for which a source size of $5''$ is assumed.  To derive the column density of this component, we use the H$_2$$^{17}$O transitions, subtract the flux predicted by the model to the $2''$ component described above, and derive the upper state column densities for each transition, assuming that the second component is optically thin.  As this may not be true, this column density should be viewed as a lower limit.  Assuming $T_\textnormal{kin} = 200$ K and $n$(H$_2$) = $10^8$ cm$^{-3}$, a correction factor of 2.1 is calculated for \emph{ortho}-H$_2$$^{17}$O, and 2.3 for \emph{para}-H$_2$$^{17}$O, which yields a H$_2$$^{17}$O column density of $2.5 \times 10^{15}$ cm$^{-2}$.  This implies a H$_2$$^{16}$O column density of $2.3 \times 10^{18}$ cm$^{-2}$.

\subsubsection{D$_2$O}

The D$_2$O isotopologue was first identified in the interstellar medium in IRAS 16293-2422 \citep{butner07, vastel10}.  Six transitions of D$_2$O were detected in Orion KL in the HIFI survey, with an average $v_\textnormal{LSR} = 7.5$ km s$^{-1}$ and $\Delta v = 4.3$ km s$^{-1}$.  As with the HD$^{18}$O lines, these are anticipated to be the most emissive transitions of D$_2$O in the HIFI bandwidth (neglecting lines that are not detected due to blends with stronger transitions of other molecules).  The kinematic parameters of the D$_2$O transitions are slightly different from those for HD$^{18}$O or the high-energy H$_2$$^{18}$O transitions, but the differences are small enough that we consider the most likely possibility to be that these components are mostly cospatial.  Collisional excitation rates for D$_2$O were recently published \citep{faure12} but extend up to only $T_\textnormal{kin} = 100$ K and are available only for low-lying energy levels, so instead we model this molecule with a LTE rotation diagram analysis \citep{goldsmith99}, assuming a statistical \emph{ortho}:\emph{para} ratio of 2:1 (different from the 3:1 of H$_2$O because of the difference between hydrogen and deuterium spin statistics).  Assuming all lines are optically thin, a rotational temperature of $74 \pm 27$ K is derived.  A similar analysis of the six detected transitions of HD$^{18}$O yields a rotational temperature of $T_\textnormal{rot} = 104 \pm 14$ K, in statistical agreement with the temperature derived for D$_2$O.  Assuming the same $2''$ source size as for the high-abundance ($v_\textnormal{LSR} \sim 7$ km s$^{-1}$) H$_2$O and HDO component, we derive $N$(D$_2$O) = $(9.6 \pm 5.5) \times 10^{14}$ cm$^{-2}$.  This results in a value of [D$_2$O]/[HDO] = $0.0016 \pm 0.0013$ in this component.

\subsection{Compact ridge}

The Compact Ridge component, as Figures 2 and 3 show, appears as a narrow spike in the line profile.  As indicated in Table 1, the Compact Ridge has generally been found to be cooler and less dense than the Hot Core \citep{blake87, tercero10}.  Figure 6 shows that HDO emission in the 8--11 km s$^{-1}$ velocity range arises from both the Compact Ridge region and a clump to the northwest (MF4/MF5 in the nomenclature of \cite{favre11}).  However, for this analysis we treat this spectral component as a single homogeneous one with a diameter of $6''$, based on the spatial extent of the HDO emission in the Compact Ridge velocity range in the ALMA images.  This component appears only in the lower-energy lines ($E_\textnormal{up} < 310$ K), indicating that the molecular gas in the Compact Ridge is less excited than in the Hot Core.  Additionally, this component is not detected in most of the lines in bands 6--7 (where the noise level is highest).  Therefore, particularly for H$_2$O, the population summation method cannot be used reliably:  combining H$_2$$^{18}$O and H$_2$$^{17}$O, the 10 transitions with a detected Compact Ridge component (5 for each isotopologue) include only 6 upper-state energy levels, 3 of \emph{ortho} and 3 of \emph{para}.  For HDO, a total of 14 transitions have detected Compact Ridge components, giving information on the population in 11 energy levels.  Therefore, we derive the physical parameters and abundances for H$_2$O and HDO in this region using RADEX models.

There are three free parameters in the modeling:  the kinetic temperature, the H$_2$ density, and the HDO or H$_2$O column density.  We assume the observed continuum field presented in Figure 5.  The figure of merit for these models was the reduced chi-squared statistic, given by

\begin{equation}
\chi^2_\textnormal{red} = \frac{1}{f} \displaystyle\sum\limits_{i=1}^n \left(\frac{W_{i,\textnormal{calc}} - W_{i,\textnormal{obs}}}{\sigma_i}\right)^2
\end{equation}

\noindent where \emph{f} is the number of degrees of freedom in the model, and $\sigma$ is the uncertainty in the line integrated flux.  The best fit models to H$_2$O and HDO emission from the Compact Ridge are presented in Figure 9.  The optimal excitation parameters are $T_\textnormal{kin} = 125$ K and $n$(H$_2$) = $10^7$ cm$^{-3}$.  An \emph{ortho}:\emph{para} ratio of 3 is assumed for the H$_2$$^{18}$O and H$_2$$^{17}$O models, and adoping a lower ratio than 3 significantly worsens the fit (as optical depths are lower than in the Hot Core lines in Figure 7).  If the enhanced continuum in Figure 5 is used instead of the observed continuum, the fit is significantly worsened, suggesting that the infrared excitation field in the Compact Ridge is lower than in the Hot Core.

\begin{figure}
\begin{center}
\includegraphics[width=5.0in]{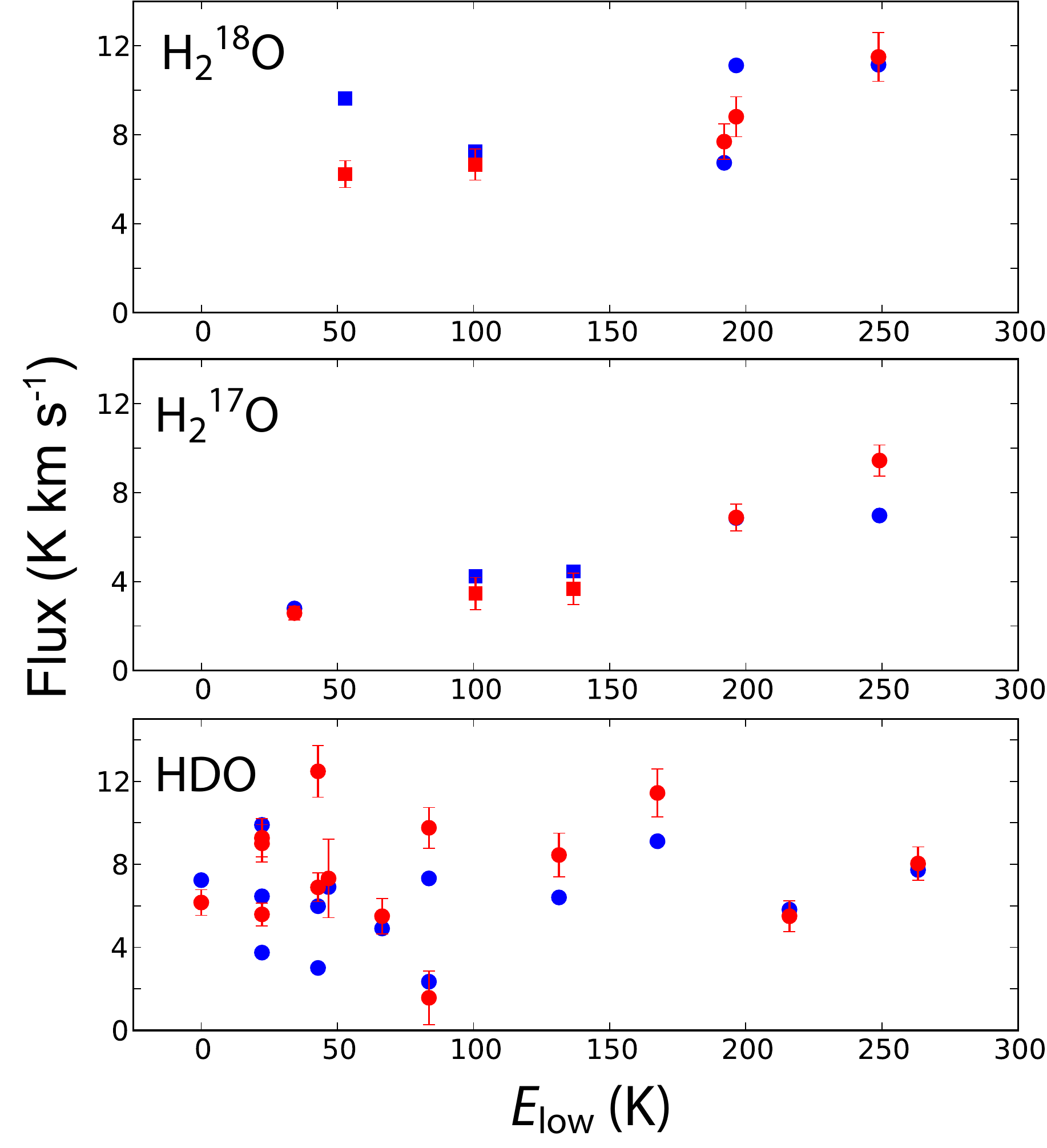}
\caption{Single-component RADEX models of H$_2$$^{18}$O, H$_2$$^{17}$O, and HDO emission in the Compact Ridge.  The common parameters to all three models are: $T_{kin}$ = 125 K, $n$(H$_2$) = $10^7$ cm$^{-3}$, $\theta_s = 6''$, and the observed continuum field from Figure 5.  The molecular column densities are $N$(H$_2$$^{18}$O) = $4.1 \times 10^{15}$ cm$^{-2}$ and $N$(HDO) = $3.9 \times 10^{15}$ cm$^{-2}$.  The red points indicate the observed line fluxes, while the blue points indicate the model fluxes.  For H$_2$$^{18}$O and H$_2$$^{17}$O, circles indicate \emph{ortho} transitions, while squares incdicate \emph{para}.  An \emph{ortho}:\emph{para} ratio of 3:1 is assumed for H$_2$$^{18}$O and H$_2$$^{17}$O.}
\end{center}
\end{figure}

\subsection{Plateau}

The emission and absorption components of the outflow are treated separately in this work.  The emissive Plateau component makes up most of the integrated flux in the lower-energy lines (see Figure 2 and 3), and is detected in nearly all lines up to $E_\textnormal{up} = 500$ K.  For this component, we use a source size of $30''$ based on a HIFI map of the H$_2$$^{16}$O $2_{12}-1_{01}$ transition (Melnick et al., in preparation).  Because this component is detected in so many transitions, we apply the population correction method to derive the column density, estimating the optical depth by comparison of corresponding H$_2$$^{18}$O and H$_2$$^{17}$O transitions as explained above.  Many of these H$_2$$^{18}$O transitions have moderate optical depth ($\tau \sim$ 1--2).  For the transitions where one of the two isotopologues is not detected due to blends with transitions of other molecules, we assume the usable line is optically thin.  To derive a correction factor, we use $T_\textnormal{kin} = 125$ K, $n$(H$_2$) = $10^7$ cm$^{-3}$, and the observed continuum in Figure 5.  This yields $f_c = 2.25$ for \emph{ortho}-water and 1.80 for \emph{para}-water, and so derive a total column density $N$(H$_2$$^{18}$O) of $(3.5 \pm 0.6) \times 10^{15}$ cm$^{-2}$, and an \emph{ortho}:\emph{para} ratio of $2.27 \pm 0.73$.  For HDO, assuming that lines are optically thin, and using the more optically thin transition in cases where two lines are detected with the same upper state, and using a correction factor of 1.6 (derived with the same parameters as for H$_2$O), a column density $N$(HDO) = $(1.23 \pm 0.25) \times 10^{15}$ cm$^{-2}$ is found.  In the error propagations, we assume a 20\% uncertainty in each correction factor.

The absorption component is detected in several low-energy lines of both HDO and H$_2$O:  two transitions ($1_{11}-0_{00}$ and $2_{12}-1_{01}$) of H$_2$$^{18}$O, H$_2$$^{17}$O, and HDO, as well as two higher-energy transitions ($2_{21}-2_{12}$ and $3_{03}-2_{12}$) of H$_2$$^{18}$O.  Fits to the eight transitions with a detected absorption component are shown in Figure 10.  The line profiles are fit to the following equation (following \cite{melnick10}):

\begin{equation}
T_{mb}(v) = (I_\textnormal{continuum} + G_\textnormal{HC}(v) + G_\textnormal{CR}(v) + G_\textnormal{PL}(v)) e^{-G_\textnormal{abs} (v)}
\end{equation}

\noindent Here, $G_\textnormal{HC}$, $G_\textnormal{CR}$, and $G_\textnormal{PL}$ are Gaussian components corresponding to the three emissive components in Orion KL with the velocity parameters given in Tables 3-5; and $G_\textnormal{abs}$ is a Gaussian component corresponding to the absorption component.  \cite{melnick10} also included in the fits to the line profiles a narrow ($\Delta v = 6.7$ km s$^{-1}$) absorption component in addition to the broad one used here, but this is only seen in H$_2$$^{16}$O transitions and not in the rare isotopologues so it is not included here.  In Figure 11 and Table 6, the intensity of the absorption component is presented as $|\Delta T_\textnormal{abs}/T_\textnormal{bg}|$, which is equal to $(1-e^{-G_\textnormal{abs}})$ at the peak of the absorption component.  We assume that both the continuum and the water absorbing layer fill the Herschel beam.  For the lines detected in HIFI bands 6 and 7, where the beamwidth is $\sim 12''$, spectra were acquired with two pointings, one near the Hot Core peak and the other near the nominal Compact Ridge, separated by $8''$.  The pointing error in these observations is estimated as $3''$.  Figure 10 shows that the absorption wing has a very similar intensity and profile in the two pointings.  This suggests that the treatment of the absorbing gas as spatially extended is reasonable.  In these fits, an LSR velocity of -5.1 km s$^{-1}$ and a width of 30 km s$^{-1}$ \citep{melnick10} is assumed, and these parameters are not varied in the fit in order to avoid a fit with too many free parameters.

\begin{figure}
\begin{center}
\includegraphics[width=3in]{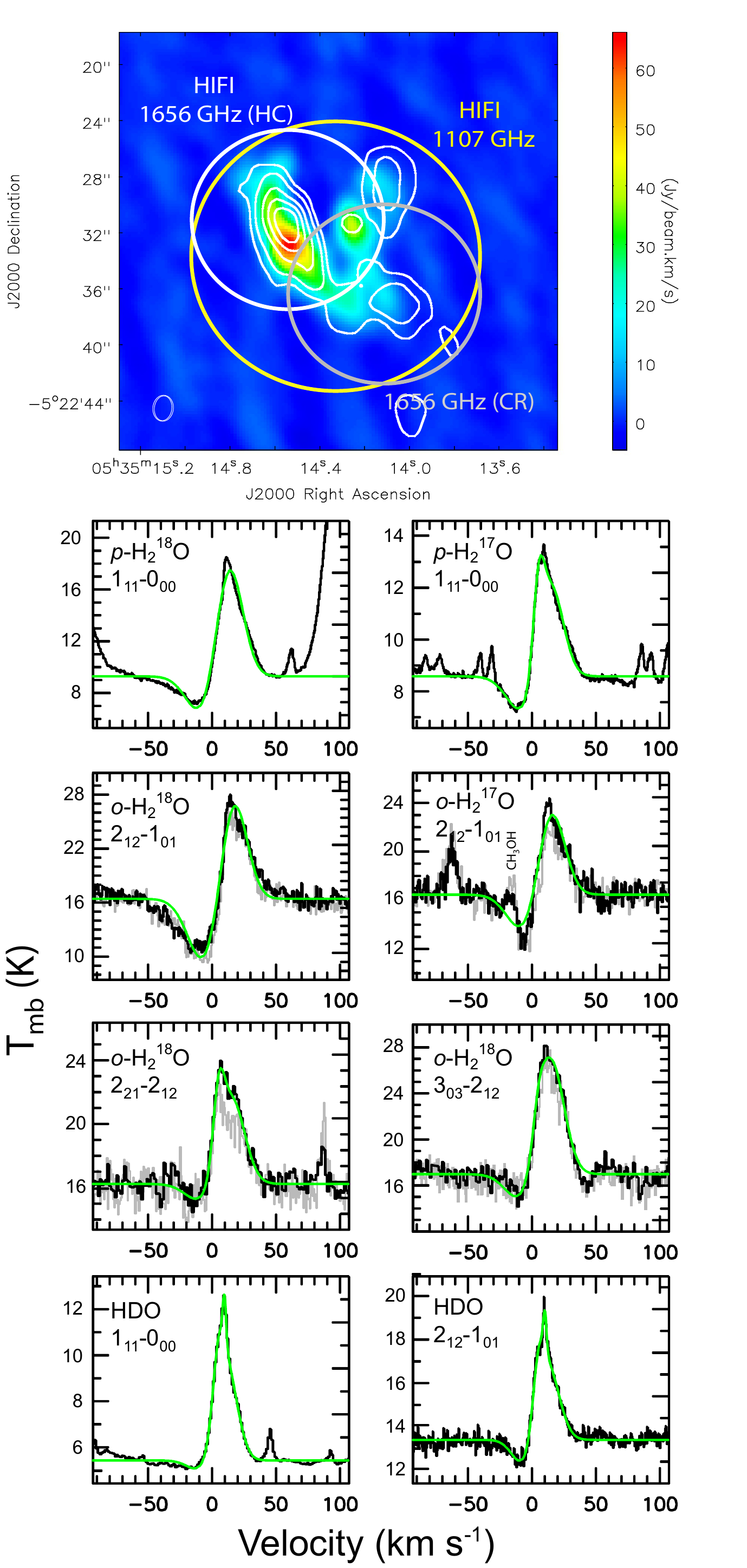}
\caption{Fits to the transitions of H$_2$$^{18}$O, H$_2$$^{17}$O, and HDO with a detected absorption component.  In the top panel, the color scale is the total integrated flux (from 2.8--11.5 km/s) of the $3_{12}-2_{21}$ transition of HDO from ALMA, the white contours are the continuum emission at 230 GHz (same contour levels as in Figure 6).  The half-power beamwidth and pointing of the HIFI Orion KL spectrum at 1107 GHz (the approximate frequency of the $1_{11}-0_{00}$ transitions of H$_2$$^{18}$O and H$_2$$^{17}$O) and both the Hot Core (HC) and Compact Ridge (CR) pointings at 1656 GHz (the frequency of the $2_{12}-1_{01}$ transitions) are overlaid.  In the lower panels, the green curves are the fits to the transitions following equation (5).  For the \emph{ortho}-H$_2$$^{18}$O and H$_2$$^{17}$O transitions, the black spectrum is the Hot Core pointing, and the gray spectrum is the Compact Ridge pointing.  The blue wing of the H$_2$$^{17}$O $2_{12}-1_{01}$ is contaminated by a transition of CH$_3$OH (the $16_6-15_5$ of the $E$ torsional subspecies at 1662586.2 MHz).  This transition is centered at a velocity of -14 km s$^{-1}$ in the reference frame of the H$_2$$^{17}$O transition.}
\end{center}
\end{figure}

Figure 11 shows model line/continuum ratios for the absorption components under a range of values for $T_\textnormal{ex}$ and column density in order to constrain the H$_2$O and HDO column density.  Panel A shows the two detected ground state ($1_{11}-0_{00}$) transitions of \emph{para}-H$_2$$^{18}$O and H$_2$$^{17}$O.  The black lines surround the parameter space where the two transitions are both fit within $1\sigma$, which yield the values $N$(\emph{p}-H$_2$$^{18}$O) = $2.8_{-1.8}^{+3.4} \times 10^{14}$ cm$^{-2}$ and $T_\textnormal{ex}$ = $23.2_{-3.4}^{+1.4}$ K.  This uncertainty also includes a 20\% error due to the uncertainty in the $^{18}$O/$^{17}$O ratio.  In these calculations, we assume that all energy levels are in LTE at the derived $T_\textnormal{ex}$.  However, the correction to the total column density located in higher energy levels (i.e., not $0_{00}$ or $1_{11}$) is likely to be small ($\sim$ 10\%), so this derivation is not extremely sensitive to non-LTE excitation.  For \emph{ortho}-H$_2$O, a similar analysis of the $2_{12}-1_{01}$ ground state transitions yields $N$(\emph{o}-H$_2$$^{18}$O) = $4.2_{-2.7}^{+3.6} \times 10^{14}$ cm$^{-2}$ and $T_\textnormal{ex}$ = $32.2_{-11.1}^{+3.3}$ K (the black lines in panel B).  Meanwhile, looking at the ground state and the two higher-energy H$_2$$^{18}$O lines (the region outlined in gray in panel B), we derive a column density of $9.8_{-1.9}^{+2.2} \times 10^{14}$ cm$^{-2}$ and $T_\textnormal{ex}$ = $34.7_{-1.5}^{+1.3}$ K.  In Figure 11, the $2_{21}-2_{12}$ transition is not plotted; its contours overlap with those of the $3_{03}-2_{12}$ transition within the uncertainties.  The average of these two analyses, $N$(\emph{ortho}-H$_2$$^{18}$O) = $7.0_{-2.3}^{+2.9} \times 10^{14}$ cm$^{-2}$, is taken as the best estimate.  Unlike for \emph{para}, there is a low-lying state ($1_{10}$) at $E = 26.3$ K above the ground state, so the correction for population in missing levels is more significant ($\sim 30\%$).  For HDO (panel C), analysis of the two transitions yields $N$(HDO) = $5.5_{-0.7}^{+0.6} \times 10^{13}$ cm$^{-2}$ and $T_\textnormal{ex}$ = $16.9_{-0.8}^{+0.8}$ K.

\begin{figure}
\begin{center}
\includegraphics[width=3.5in]{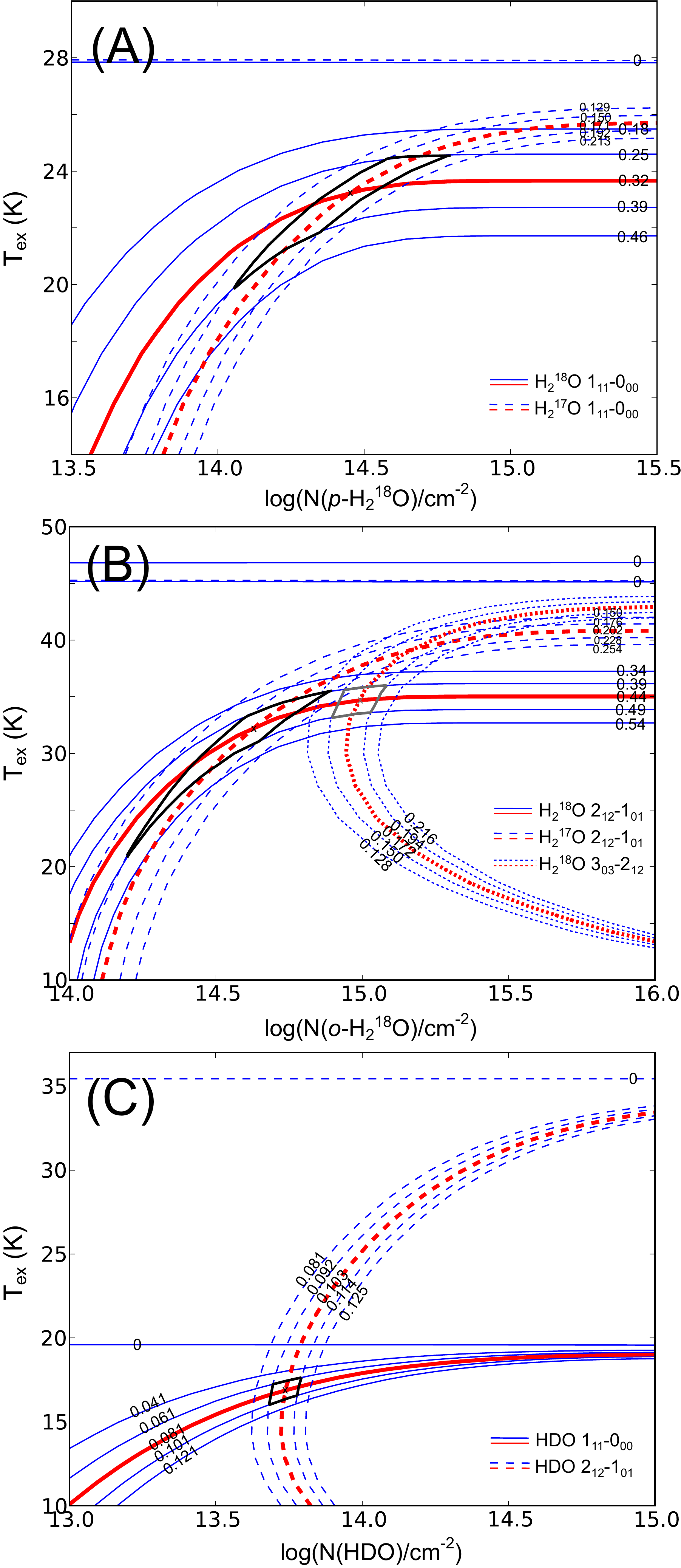}
\caption{Models to the absorption component of H$_2$O and HDO.  The red contours indicate the absolute value of the fit line/continuum ratio for each transition, while the blue contours indicate the $\pm 1\sigma$ and $\pm 2\sigma$ error bars.  The black and gray lines indicate the parameter space where the $\pm 1\sigma$ error bars intersect for two transitions.}
\end{center}
\end{figure}  

D$_2$O is not detected in either the emission or absorption components of the outflowing gas.  As Figure 4 shows, particularly the ground state transitions ($1_{11}-0_{00}$ for \emph{ortho}, $2_{12}-1_{01}$ for \emph{para}) are fairly clean in the wings where these components would be detected.  Using LTE models, we estimate an upper limit to the [D$_2$O]/[HDO] ratio of $\sim 0.1$ in both components.  A ratio of 0.01 or less would be expected based on the [HDO]/[H$_2$O] ratio in these components (Table 2).

\section{Discussion}

We note some differences between the H$_2$O abundances derived here and those of \cite{melnick10}, which were derived using the same data set.  These differences could arise either in the Gaussian fitting process (i.e., the attribution of the total flux observed by HIFI to the various spatial components) or in the derivation of the water column density from these Gaussian components.  The only component strongly affected by differences in the Gaussian fitting is the Compact Ridge, which was fit in a different way between the two analyses.  In \cite{melnick10}, the linewidth for the Gaussians attributed to this component (which was called ``extended warm gas'') was allowed to vary between 2--8 km s$^{-1}$, which likely encompasses flux belonging to regions not identified with the spatially and spectrally distinct Compact Ridge region.  This is the most significant reason for the lower Compact Ridge H$_2$O abundance derived in this study, though other methodological assumptions also contribute.  For the Hot Core and Plateau, on the other hand, the fluxes attributed to these components are similar, within 20\% for most transitions.

The Hot Core was modeled as a single component in the previous study, but with two spatial components here, which is largely responsible for the different abundances, and particularly the higher water abundance in the $2''$, 7 km s$^{-1}$ component.  For the emissive Plateau component, we derive a H$_2$O abundance a factor of 15 lower than that of \cite{melnick10}.  Some of this difference is due to factors connected with the conversion of the H$_2$$^{18}$O column density (the primary quantity derived by the radiative transfer modeling) to a H$_2$O abundance relative to H$_2$:  a solar $^{16}$O/$^{18}$O isotopic ratio of 500 was assumed by \cite{melnick10}, whereas a value of 250 is used here \citep{tercero10}.  Additionally, the H$_2$ column density used for that study was $1.0 \times 10^{23}$ cm$^{-2}$ from \cite{blake87}, while we use $1.8 \times 10^{23}$ cm$^{-2}$ \citep{plume12}.  Both of these factors lower the H$_2$O abundance from that of \cite{melnick10}; however, there is still a factor of 4 difference in the H$_2$$^{18}$O column density that is attributed to differences in the methods used for the column density derivation.  The previous analysis did not use H$_2$$^{17}$O transitions as a constraint on the optical depth of the H$_2$$^{18}$O lines.  In the analysis presented here, the column density is sensitive only to two factors:  the optical depth estimates, and the value of $f_c$ used to account for population in unprobed levels.  The optical depth estimates depend on the $^{18}$O/$^{17}$O ratio, as discussed above, and also assume that $T_\textnormal{ex}$ is the same between corresponding transitions of the two isotopologues.  RADEX modeling suggests that this may not always be the case, and small deviations (10--20\%) from this assumption can cause large uncertainties, a factor of 2 or more, in the population in individual levels.  We find that the correction factor is relatively insensitive to physical conditions, and particularly to the intensity of the radiation field; most of the unprobed population is located in the ground state levels, which are not as strongly affected by the far-IR continuum.

The component with highest water abundance is the small $v_\textnormal{LSR} = 7$ km s$^{-1}$ clump within the Hot Core region, where most of the oxygen is in gas-phase water.  The other spatial components have lower abundances by about two orders of magnitude (2.7--6.7 $\times 10^{-6}$).  This high abundance and the high [HDO]/[H$_2$O] ratio of 0.003 observed in the 7 km s$^{-1}$ Hot Core component (and a comparable [D$_2$O]/[HDO] ratio of 0.0016) suggest that much of this water is material that has been recently evaported from ice mantles.  Low-temperature gas-phase chemistry could produce water with significant deuterium fractionation, for example, through ion-neutral reactions involving H$_2$D$^+$ \citep{millar89}, but likely not with such a high water abundance \citep{herbst73, woodall07}.  Gas-phase neutral-neutral chemistry in shocked gas, on the other hand, can produce H$_2$O abundances of $\sim 10^{-4}$ \citep{draine83, bergin98}, but not with such high deuterium fractionation due to the high temperatures required \citep{bergin99}.  The more spatially extended component in the Hot Core (with velocity centered at 5 km s$^{-1}$) has a lower abundance of H$_2$O, but this value is a lower limit.  Similar deuterium fractionation is found in the two Hot Core components.

The Compact Ridge has somewhat higher deuteration than the Hot Core, suggesting that H$_2$O in this region may have been synthesized under slightly colder conditions.  The spatial distribution of HDO emission (Figure 6) has an interesting morphology, with the strongest emission found to the northeast of the continuum peak, in the part of the Compact Ridge facing nearest to the Hot Core (and so nearest to the origin of the molecular outflow), rather than where other oxygen-bearing organic species peak; e.g. the MF1 peak in Figure 5 is the region of strongest methyl formate (HCOOCH$_3$) emission \citep{favre11}.  The Compact Ridge has been suggested to be a site of recent interaction between the molecular outflow from Source I \citep{zapata12} and pre-existing dense gas, leading to the liberation of organic material from ice mantles \citep{blake87, liu02, favre11}.  However, physical conditions will also play a role in the excitation of these HDO transitions.

The emission component of the Plateau has a somewhat lower HDO/H$_2$O ratio than the compact regions, and the absorption component has lower deuterium fractionation by an order of magnitude.  This suggests that most of the water in the outflow, and particularly in the absorbing gas, does not have the same origin as the more deuterated water in the Hot Core and Compact Ridge.  The HDO/H$_2$O ratio can be modified in shocks by gas-phase neutral-neutral chemistry \citep{kaufman96, bergin98, bergin99}.  Water synthesis in shocked gas proceeds by the following mechanism:

\begin{equation}
\textnormal{O} + \textnormal{H}_2 \rightarrow \textnormal{OH} + \textnormal{H}
\end{equation}
\begin{equation}
\textnormal{OH} + \textnormal{H}_2 \rightarrow \textnormal{H}_2\textnormal{O} + \textnormal{H}
\end{equation}

\noindent HDO can be formed through similar chemistry, with either
\begin{equation}
\textnormal{O} + \textnormal{HD} \rightarrow \textnormal{OD} + \textnormal{H}
\end{equation}

\noindent or

\begin{equation}
\textnormal{OH} + \textnormal{HD} \rightarrow \textnormal{HDO} + \textnormal{H}
\end{equation}

\noindent in place of Eq. (5) or (6).  Rates for the relevant reactions are available through the UMIST astrochemistry database \citep{woodall07} and in \cite{bergin99}.  These reactions, particularly equations (5) and (7), have substantial energy barriers (e.g., 3160 K for equation (5)).  However, in a sufficiently energetic shock, this set of reactions can nevertheless convert all oxygen not in CO into water.  For example, in a C-type shock with a velocity of 20 km s$^{-1}$, corresponding to a peak gas temperature of 1000 K \citep{kaufman96}, and with an H$_2$ density of $10^5$ cm$^{-3}$, the pseudo-first order reaction rate of equation (5) is $3.5 \times 10^{-8}$ s$^{-1}$.  This corresponds to a timescale for the conversion of O to H$_2$O of 0.9 yr, far shorter than the lifetime of the shock \citep{bergin98}, so atomic oxygen will be readily converted to water by these reactions.  Equations (7) and (8) have far slower rates, due to the low abundance of HD relative to H$_2$ ($\sim 2 \times 10^{-5}$); under these same conditions, equation (7) has a pseudo-first order rate of $2.3 \times 10^{-13}$ s$^{-1}$.  A kinetic analysis shows that a 20 km s$^{-1}$ C-type shock ($T_\textnormal{gas} = 1000$ K) with an H$_2$ density of $10^5$ cm$^{-3}$ and [HD]/[H$_2$] = $2 \times 10^{-5}$ will produce water with [HDO]/[H$_2$O] = $8 \times 10^{-6}$.  This is lower than the [HD]/[H$_2$] ratio due to slower rate constants for the reactions involving deuterium.

The [HDO]/[H$_2$O] ratios observed both the emission and absorption components of the outflow are intermediate between the ratio observed in the Hot Core and Compact Ridge and the low ratio anticipated by the shock gas-phase water production mechanism.  This suggests that the gas in the outflow originated close to the core of Orion KL with a higher HDO/H$_2$O ratio, possibly similar to the fractionation observed in the quiescent components, and the fractionation has been reduced by the production of additional water in the outflow with a low HDO/H$_2$O ratio.  OH$^+$ and H$_2$O$^+$ have been detected in the absorbing layer \citep{gupta10}; it was proposed by these authors that these ions could be produced by the photodissociation of H$_2$O.  If water has a substantial destruction rate in the outflow, the original water that originated from ice mantles is destroyed on a relatively short timescale and could be replaced with fresh water with a low HDO/H$_2$O ratio produced via high-temperature gas-phase chemistry.  The deuterium fractionation in outflows could therefore reflect both the physical conditions in the preshocked gas and in the shock itself.  Additionally, we note that the H$_2$O abundance in the outflow ($\sim 3 \times 10^{-6}$) is low compared to the ISO studies of \cite{harwit98} and \cite{cernicharo06}, who find beam-averaged H$_2$O abundances of $2\times 10^{-5}-4 \times 10^{-4}$.  These analyses were primarily concerned with transitions of the H$_2$$^{16}$O isotopologue, which are significantly broader than those of H$_2$$^{18}$O and H$_2$$^{17}$O, also seen in \cite{melnick10}, and therefore are probing more of the high-velocity shocks.  The present analysis, focusing on rare isotopologues of water, is concerned with the regions of highest water optical depth closer to the KL nebula, which may have lower water abundance relative to H$_2$ than the faster shocks.

The factor of 6 difference in the D/H ratios between the emission and absorption components of the outflow is intriguing, and significant within the errors in our analysis.  For the emission component, the optical depth of the H$_2$O lines is likely well characterized due to the detection of both H$_2$$^{18}$O and H$_2$$^{17}$O, although violation of the assumption that $T_\textnormal{ex}$ is the same for corresponding H$_2$$^{18}$O and H$_2$$^{17}$O transitions may add uncertainty.  Line optical depths are less well chracterized for HDO, but if the opacity has been underestimated the effect will be to increase, rather than decrease, the D/H ratio in the emitting gas.  This suggests a chemical difference between the emissive gas of the Plateau and the absorbing layer.

\section{Conclusion}

Using the HIFI fullband survey of Orion KL, acquired as part of the HEXOS key program, we have detected numerous transitions of isotopologues of H$_2$O (H$_2$$^{18}$O, H$_2$$^{17}$O, HDO, HD$^{18}$O, and D$_2$O) with a variety of excitation conditions.  We have derived abundances of H$_2$O and HDO in each of the spatial components within this region.  Water has a complex morphology in Orion KL, with significant H$_2$O and HDO emission in the Hot Core, Compact Ridge, and Plateau, as well as absorption in the blue-shifted wing of the outflow in a few low-energy transitions.  Both the H$_2$O abundance and HDO/H$_2$O ratio have significant differences between spatial components, and we propose some possible explanations for these variations.  Of particular interest is the small ($2''$) clump we identify in the Hot Core region, which we attribute to a region just south of the dust continuum peak, and near (but not coincident with) the SMA1 submillimeter continuum peak of \cite{beuther04}.  This region has a very high abundance of water, with a high [HDO]/[H$_2$O] ratio (0.003), suggesting material that was formed at low temperatures and has been recently evaporated from ice mantles.  This region also shows signs of significant excitation from a nearby far-IR field, possibly from an embedded far-IR continuum source, in agreement with the recent study of H$_2$S in the Hot Core (Crockett et al. 2013a, in preparation).  The far-IR dust opacity is likely to be very high in this region, which make continuum sources difficult to detect directly.  Further investigations into the spatial distributions of transitions of molecules that trace the far-IR radiation field will be crucial in investigating the physical structure of this region.

\begin{acknowledgements}

HIFI has been designed and built by a consortium of institutes and university departments from across Europe, Canada, and the United States under the leadership of SRON Netherlands Institute for Space Research, Groningen, The Netherlands and with major contributions from Germany, France, and the US.  Consortium members are: Canada: CSA, U.Waterloo; France: CESR, LAB, LERMA, IRAM; Germany: KOSMA, MPIfR, MPS; Ireland: NUI Maynooth; Italy: ASI, IFSI-INAF, Osservatorio Astrofisico di Arcetri-INAF; Netherlands: SRON, TUD; Poland: CAMK, CBK; Spain: Observatorio Astron\'{o}mico Nacional (IGN), Centro de Astrobiolog\'{i}a (CSIC-INTA); Sweden: Chalmers University of Technology--MC2, RSS \& GARD, Onsala Space Observatory, Swedish National Space Board, Stockholm Observatory; Switzerland: ETH Zurich, FHNW; USA: Caltech, JPL, NHSC.  Support for this work was provided by NASA through an award issued by JPL/Caltech.

This paper makes use of the following ALMA data:  ADS/JAO.ALMA\#2011.0.00009.SV.  ALMA is a partnership of ESO (representing its member states), NSF (USA) and NINS (Japan), together with NRC (Canada) and NSC and ASIAA (Taiwan), in cooperation with the Republic of Chile.  The Joint ALMA Observatory is operated by ESO, AUI/NRAO and NAOJ.

We thank the anonymous referee for carefully reading and offering comments to improve the manuscript. 

\end{acknowledgements}

{\it Facilities:} \facility{Herschel}, \facility{ALMA}

\bibliography{h2obib}

\clearpage
\appendix
\section{Fit line parameters for H$_2$O isotopologues}
\begin{deluxetable}{c c c c c c c}
\tablewidth{0pt}
\tablecaption{Fit integrated fluxes for H$_2$O isotopologues in the Hot Core.}
\tablehead{
Transition & Frequency & $E_u$ & $S_\textnormal{ij}\mu^2$ & $\int T_\textnormal{mb}dv$ & $v_\textnormal{LSR}$\tablenotemark{a} & $\Delta v$\tablenotemark{a} \\
 & (MHz) & (K) & (D$^2$) & (K km s$^{-1}$) & (km s$^{-1}$) & (km s$^{-1}$)}
\startdata
H$_2$$^{18}$O \\
\hline
$2_{02}-1_{11}$ & 994675.1 & 100.6 & 2.63 & 48.7(5.7) & 5.2 & 10.0 \\
$2_{11}-2_{02}$ & 745320.2 & 136.4 & 7.09 & 37.1(4.3) & 5.2 & 10.0 \\
$2_{21}-2_{12}$ & 1633483.6 & 192.0 & 8.60 & 40.2(6.5) & 5.2 & 10.0 \\
$3_{03}-2_{12}$ & 1719250.2 & 196.2 & 18.16 & 22.4(6.1) & 5.2 & 10.0 \\
$3_{12}-2_{21}$ & 1181394.0 & 248.7 & 3.17 & 62.2(6.4) & 3.6(0.1) & 8.9(0.1) \\
$3_{12}-3_{03}$ & 1095627.4 & 248.7 & 22.24 & 56.1(6.8) & 3.6(0.2) & 9.6(0.4) \\
$3_{22}-3_{13}$ & 1894323.8 & 294.6 & 4.45 & 49.3(7.0) & 4.9(0.2) & 8.1(0.6) \\
$3_{21}-3_{12}$ & 1136703.6 & 303.3 & 26.42 & 56.9(5.7) & 5.2 & 10.0 \\
$4_{13}-4_{04}$ & 1605962.5 & 395.4 & 6.93 & 38.3(7.6) & 4.87(0.2) & 6.4(0.8) \\
$4_{22}-4_{13}$ & 1188863.1 & 452.4 & 12.55 & 80.6(8.9) & 5.2 & 10.0 \\
$5_{24}-4_{31}$ & 1003277.6 & 595.9 & 0.90 & 9.0(1.0) & 5.6(0.1) & 6.7(0.3) \\
$5_{32}-4_{41}$ & 692079.1 & 727.6 & 1.26 & 4.3(0.5) & 5.2 & 8.2(0.4) \\
$5_{32}-5_{23}$ & 1815853.4 & 727.6 & 35.87 & 70.7(7.5) & 5.8(0.2) & 9.0(0.4) \\
$6_{24}-6_{15}$ & 1800474.6 & 865.0 & 14.35 & 30.0(3.5) & 3.5(0.2) & 7.2(0.5) \\
$6_{34}-5_{41}$ & 1216850.4 & 928.6 & 2.92 & 8.8(1.0) & 5.3(0.3) & 10.0 \\
$6_{33}-5_{42}$ & 1620851.6 & 947.6 & 1.08 & 27.6(3.9) & 4.1(0.5) & 10.4(1.3) \\
$7_{34}-7_{25}$ & 1771674.6 & 1207.9 & 61.01 & 14.9(3.4) & 5.7(0.6) & 6.7(1.8) \\
\hline
H$_2$$^{17}$O \\
\hline
$1_{11}-0_{00}$ & 1107166.9 & 53.1 & 3.44 & 29.7(3.8) & 5.2 & 10.0 \\
$1_{10}-1_{01}$ & 552021.0 & 60.7 & 15.48 & 6.6(0.9) & 5.2 & 10.0 \\
$2_{11}-2_{02}$ & 748458.3 & 136.6 & 7.11 & 37.3(17.6) & 5.2 & 10.0 \\
$2_{21}-2_{12}$ & 1646398.7 & 193.0 & 8.60 & 27.5(6.3) & 5.2 & 10.0 \\
$2_{20}-2_{11}$ & 1212980.4 & 194.9 & 4.36 & 26.2(3.9) & 5.2 & 10.0 \\
$3_{03}-2_{12}$ & 1718119.5 & 196.5 & 18.08 & 54.5(9.1) & 5.2 & 10.0 \\
$3_{12}-3_{03}$ & 1096414.3 & 249.1 & 22.38 & 57.6(5.8) & 3.7(0.1) & 10.0 \\
$3_{21}-3_{12}$ & 1148976.1 & 304.2 & 26.31 & 57.5(6.2) & 4.2(0.2) & 10.0 \\
$4_{13}-4_{04}$ & 1604179.9 & 395.9 & 6.98 & 42.0(11.9) & 5.3(0.3) & 8.8(1.3) \\
$4_{22}-4_{13}$ & 1197610.3 & 453.3 & 12.54 & 33.8(6.8) & 5.1(0.7) & 7.9(2.3) \\
$5_{32}-5_{23}$ & 1840155.7 & 729.7 & 35.63 & 32.2(4.0) & 5.7(0.3) & 7.0(0.6) \\
$6_{24}-6_{15}$ & 1797675.5 & 866.1 & 14.47 & 16.0(4.3) & 5.2(0.7) & 7.5(2.2) \\
$7_{34}-7_{25}$ & 1783388.8 & 1209.8 & 60.99 & 9.6(3.5) & 7.2(1.2) & 7.2(2.9) \\
\hline
HDO \\
\hline
$1_{11}-0_{00}$ & 893638.7 & 42.9 & 3.0 & 23.8(3.4) & 5.2 & 10.0 \\
$1_{10}-1_{01}$ & 509292.4 & 46.8 & 4.52 & 11.4(2.8) & 5.2 & 10.0 \\
$2_{02}-1_{01}$ & 919310.9 & 66.4 & 0.86 & 44.6(4.5) & 5.2 & 10.0 \\
$2_{02}-1_{11}$ & 490596.6 & 66.4 & 1.91 & 14.3(1.5) & 5.0 & 10.0 \\
$2_{12}-1_{01}$ & 1277675.9 & 83.6 & 4.53 & 27.7(2.9) & 5.2 & 10.0 \\
$2_{12}-1_{11}$ & 848961.8 & 83.6 & 0.65 & 24.5(2.5) & 5.2 & 10.0 \\
$2_{11}-1_{10}$ & 1009944.7 & 95.2 & 0.65 & 45.9(7.0) & 5.2 & 10.0 \\
$2_{11}-2_{02}$ & 599926.7 & 95.2 & 6.87 & 30.0(3.1) & 6.5(0.5) & 11.3(0.5) \\
$3_{03}-2_{12}$ & 995411.5 & 131.4 & 4.30 & 43.6(4.5) & 5.2 & 10.0 \\
$3_{13}-2_{02}$ & 1625408.1 & 144.4 & 6.29 & 24.1(7.5) & 6.6(0.8) & 10.0 \\
$2_{21}-2_{12}$ & 1522925.8 & 156.7 & 2.51 & 18.6(3.1) & 2.9(0.8) & 10.0 \\
$3_{12}-2_{11}$ & 1507261.0 & 167.6 & 1.16 & 35.6(4.7) & 4.3(0.4) & 10.0 \\
$3_{12}-3_{03}$ & 753411.2 & 167.6 & 8.30 & 39.0(5.7) & 5.2 & 10.0 \\
$4_{04}-3_{13}$ & 1491926.9 & 216.0 & 7.15 & 22.9(4.4) & 5.2 & 10.0 \\
$3_{22}-3_{13}$ & 1648801.4 & 223.6 & 4.13 & 50.3(5.7) & 5.2 & 10.0 \\
$4_{14}-3_{13}$ & 1678577.8 & 225.0 & 1.62 & 29.4(5.7) & 4.9(0.3) & 7.1(0.9) \\
$3_{21}-2_{20}$ & 1432876.7 & 226.0 & 0.72 & 45.5(9.9) & 6.3(1.0) & 10.2(2.4) \\ 
$3_{21}-3_{12}$ & 1217258.3 & 226.0 & 6.33 & 48.1(9.0) & 4.5(0.3) & 10.6(0.7) \\
$4_{13}-4_{04}$ & 984137.8 & 263.3 & 8.75 & 56.4(5.7) & 5.3(0.3) & 10.0 \\
$4_{13}-3_{22}$ & 827263.4 & 263.3 & 1.75 & 28.5(3.0) & 5.8(0.1) & 8.4(0.2) \\
$4_{23}-3_{22}$ & 1848306.0 & 312.3 & 1.30 & 40.7(4.7) & 4.4(0.2) & 8.5(0.6) \\
$4_{23}-4_{14}$ & 1818529.7 & 312.3 & 5.30 & 32.9(7.6) & 4.8(0.3) & 9.1(1.2) \\
$4_{22}-4_{13}$ & 1164769.9 & 319.2 & 9.81 & 63.7(6.8) & 5.2 & 10.0 \\
$5_{14}-4_{23}$ & 1444829.0 & 381.6 & 3.21 & 26.9(7.5) & 3.7(0.9) & 6.9(2.1) \\
$5_{14}-5_{15}$ & 1180323.5 & 381.6 & 0.17 & 11.5(1.2) & 5.8(1.3) & 6.8(0.4) \\
$4_{32}-3_{31}$ & 1872608.6 & 425.1 & 0.76 & 26.7(3.4) & 5.4(0.3) & 7.5(0.7) \\
$4_{31}-3_{30}$ & 1877486.8 & 425.4 & 0.76 & 18.0(2.9) & 5.6(0.4) & 6.3(1.0) \\
$6_{15}-6_{06}$ & 1684605.8 & 521.6 & 8.12 & 56.5(6.8) & 4.7(0.3) & 12.0(1.2) \\
$6_{24}-6_{15}$ & 1230402.9 & 580.6 & 15.82 & 47.0(5.0) & 5.2(0.1) & 8.1(0.2) \\
$6_{24}-5_{33}$ & 895874.4 & 580.6 & 1.78 & 12.1(1.2) & 5.9(0.1) & 6.6(0.1) \\
$7_{26}-6_{33}$ & 622482.6 & 705.6 & 1.65 & 2.3(0.6) & 5.9 & 5.4(1.0) \\
$7_{25}-6_{34}$ & 1577177.6 & 748.3 & 2.74 & 19.3(3.9) & 6.8(0.6) & 6.3(1.2) \\
$7_{34}-7_{25}$ & 1853872.8 & 837.3 & 14.83 & 21.6(2.8) & 5.3(0.3) & 6.6(0.7) \\
$8_{27}-7_{34}$ & 838953.3 & 877.6 & 1.71 & 3.8(1.5) & 5.8(1.0) & 5.4(2.6) \\
$8_{26}-8_{17}$ & 1634639.2 & 939.6 & 17.45 & 16.3(2.9) & 6.1(0.5) & 5.9(1.1) \\
$8_{35}-8_{26}$ & 1759978.4 & 1024.1 & 18.77 & 24.5(4.9) & 4.6(0.5) & 6.5(1.3) \\
$9_{36}-9_{27}$ & 1731255.8 & 1236.5 & 22.42 & 7.4(1.4) & 6.2(0.3) & 3.1(0.6) \\
\hline
HD$^{18}$O \\
\hline
$1_{11}-0_{00}$ & 883189.4 & 42.4 & 2.98 & 3.0(0.4) & 6.6(0.2) & 5.7(0.4) \\
$2_{02}-1_{11}$ & 492814.5 & 66.0 & 1.89 & 0.4(0.2) & 6.6(1.0) & 4.7(1.0) \\
$2_{11}-2_{02}$ & 592405.7 & 94.5 & 6.78 & 2.1(0.4) & 7.0(0.6) & 6.9(1.4) \\
$3_{03}-2_{12}$ & 994348.0 & 130.6 & 4.27 & 4.9(0.9) & 7.0(1.0) & 6.9(1.4) \\
$3_{12}-3_{03}$ & 746475.6 & 166.4 & 8.16 & 1.5(0.3) & 6.3(1.3) & 4.4(1.1) \\
$4_{22}-4_{13}$ & 1144046.2 & 316.5 & 9.74 & 2.1(0.5) & 6.9(0.5) & 4.6(0.4) \\
\hline
D$_2$O \\
\hline
$1_{11}-0_{00}$ & 607349.5 & 29.1 & 6.81 & 1.08(0.12) & 7.6(0.1) & 5.5(0.3) \\
$2_{12}-1_{01}$ & 897947.1 & 60.5 & 5.11 & 3.3(0.4) & 7.7(0.1) & 4.6(0.3) \\
$2_{20}-2_{11}$ & 743563.4 & 106.7 & 7.97 & 1.02(0.18) & 7.0(0.4) & 4.8(0.8) \\
$3_{13}-2_{02}$ & 1158044.9 & 107.2 & 14.48 & 1.54(0.24) & 6.1(0.2) & 3.2(0.4) \\
$3_{21}-3_{12}$ & 697922.7 & 161.5 & 8.07 & 1.6(0.5) & 8.7(0.6) & 4.8(1.7) \\
$4_{13}-4_{04}$ & 782470.9 & 203.0 & 16.02 & 0.60(0.07) & 8.0(0.1) & 3.2(0.3)
\enddata
\tablenotetext{a}{Numbers without uncertainties indicate values that were not varied in the fit.}
\end{deluxetable}

\begin{deluxetable}{c c c c c c c}
\tablewidth{0pt}
\tablecaption{Fit integrated fluxes for H$_2$O isotopologues in the Compact Ridge.}
\tablehead{
Transition & Frequency & $E_u$ & $S_\textnormal{ij}\mu^2$ & $\int T_\textnormal{mb}dv$ & $v_\textnormal{LSR}$\tablenotemark{a} & $\Delta v$\tablenotemark{a} \\
 & (MHz) & (K) & (D$^2$) & (K km s$^{-1}$) & (km s$^{-1}$) & (km s$^{-1}$)}
\startdata
H$_2$$^{18}$O \\
\hline
$2_{02}-1_{11}$ & 994675.1 & 100.6 & 2.63 & 6.2(1.1) & 8.0 & 3.0 \\
$2_{11}-2_{02}$ & 745320.2 & 136.4 & 7.09 & 6.7(0.9) & 8.0 & 3.0 \\
$3_{12}-2_{21}$ & 1181394.0 & 248.7 & 3.17 & 7.7(0.9) & 7.6(0.1) & 3.0 \\
$3_{12}-3_{03}$ & 1095627.4 & 248.7 & 22.24 & 8.8(1.3) & 7.7(0.1) & 3.0 \\
$3_{21}-3_{12}$ & 1136703.6 & 303.3 & 26.42 & 11.5(1.2) & 8.0 & 3.0 \\
\hline
H$_2$$^{17}$O \\
\hline
$1_{10}-1_{01}$ & 552021.0 & 60.7 & 15.48 & 2.6(0.3) & 9.2(1.0) & 3.0 \\
$2_{11}-2_{02}$ & 748458.3 & 136.6 & 7.11 & 3.5(1.0) & 7.7(0.8) & 2.6(0.4) \\
$2_{20}-2_{11}$ & 1212980.4 & 194.9 & 4.36 & 3.6(1.0) & 9.0 & 3.0 \\
$3_{12}-3_{03}$ & 1096414.3 & 249.1 & 22.38 & 6.9(0.8) & 7.6(0.8) & 3.0 \\
$3_{21}-3_{12}$ & 1148976.1 & 304.2 & 26.31 & 9.4(1.4) & 8.2(0.8) & 3.0 \\
\hline
HDO \\
\hline
$1_{11}-0_{00}$ & 893638.7 & 42.9 & 3.0 & 6.2(0.8) & 9.6(0.1) & 3.0 \\
$1_{10}-1_{01}$ & 509292.4 & 46.8 & 4.52 & 5.6(0.8) & 9.4(0.2) & 3.0 \\
$2_{02}-1_{01}$ & 919310.9 & 66.4 & 0.86 & 9.0(0.9) & 8.6(0.1) & 3.0 \\
$2_{02}-1_{11}$ & 490596.6 & 66.4 & 1.91 & 12.4(1.3) & 8.6(0.1) & 3.0 \\
$2_{12}-1_{01}$ & 1277675.9 & 83.6 & 4.53 & 9.3(1.3) & 10.2(0.5) & 3.0 \\
$2_{12}-1_{11}$ & 848961.8 & 83.6 & 0.65 & 6.9(0.8) & 8.2(0.1) & 3.0 \\
$2_{11}-1_{10}$ & 1009944.7 & 95.2 & 0.65 & 7.3(2.0) & 7.9(0.4) & 3.0 \\
$2_{11}-2_{02}$ & 599926.7 & 95.2 & 6.87 & 5.5(1.0) & 8.7(0.5) & 3.0 \\
$3_{03}-2_{12}$ & 995411.5 & 131.4 & 4.30 & 9.8(1.0) & 8.2(0.1) & 3.0 \\
$2_{21}-2_{12}$ & 1522925.8 & 156.7 & 2.51 & 1.6(0.5) & 8.0 & 3.0 \\
$3_{12}-3_{03}$ & 753411.2 & 167.6 & 8.30 & 8.4(1.3) & 8.0 & 3.0 \\
$3_{21}-3_{12}$ & 1217258.3 & 226.0 & 6.33 & 11.4(1.4) & 8.0 & 3.0 \\
$4_{13}-4_{04}$ & 984137.8 & 263.3 & 8.75 & 5.5(0.9) & 8.1(0.3) & 3.0 \\
$4_{22}-4_{13}$ & 1164769.9 & 319.2 & 9.81 & 8.0(1.0) & 7.9(0.1) & 2.2(0.1)
\enddata
\tablenotetext{a}{Numbers without uncertainties indicate values that were not varied in the fit.}
\end{deluxetable}

\clearpage

\begin{deluxetable}{c c c c c c c}
\tablewidth{0pt}
\tablecaption{Fit integrated fluxes for H$_2$O isotopologues in the Plateau (emission component).}
\tablehead{
Transition & Frequency & $E_u$ & $S_\textnormal{ij}\mu^2$ & $\int T_\textnormal{mb}dv$ & $v_\textnormal{LSR}$\tablenotemark{a} & $\Delta v$\tablenotemark{a} \\
 & (MHz) & (K) & (D$^2$) & (K km s$^{-1}$) & (km s$^{-1}$) & (km s$^{-1}$)}
\startdata
H$_2$$^{18}$O \\
\hline
$1_{11}-0_{00}$ & 1101698.3 & 52.9 & 3.44 & 288(64) & 12.1(0.1) & 24.9(0.2) \\
$1_{10}-1_{01}$ & 547676.4 & 60.5 & 15.49 & 169(17) & 12.2(0.2) & 25.4(0.5) \\
$2_{02}-1_{11}$ & 994675.1 & 100.6 & 2.63 & 380(38) & 11.4(0.1) & 28.7(0.3) \\
$2_{12}-1_{01}$ & 1655867.6 & 113.7 & 15.49 & 405(41) & 14.0(0.2) & 27.0(0.3) \\
$2_{11}-2_{02}$ & 745320.2 & 136.4 & 7.09 & 272(27) & 10.6(0.2) & 27.3(0.2) \\
$2_{21}-2_{12}$ & 1633483.6 & 192.0 & 8.60 & 161.4(16.6) & 13.8(0.4) & 25.0(0.6) \\
$3_{03}-2_{12}$ & 1719250.2 & 196.2 & 18.16 & 271.5(27.6) & 14.1(0.3) & 22.6(0.4) \\
$3_{12}-2_{21}$ & 1181394.0 & 248.7 & 3.17 & 257(26) & 9.2(0.1) & 25.8(0.1) \\
$3_{12}-3_{03}$ & 1095627.4 & 248.7 & 22.24 & 436(44) & 10.8(0.1) & 27.5(0.2) \\
$3_{22}-3_{13}$ & 1894323.8 & 294.6 & 4.45 & 66.7(9.0) & 11.9(1.0) & 25.0 \\
$3_{21}-3_{12}$ & 1136703.6 & 303.3 & 26.42 & 494(49) & 10.8(0.1) & 28.5(0.1) \\
$4_{13}-4_{04}$ & 1605962.5 & 395.4 & 6.93 & 73.3(10.5) & 8.2(1.1) & 23.4(1.8) \\
$4_{22}-4_{13}$ & 1188863.1 & 452.4 & 12.55 & 159.8(16.5) & 6.7(0.2) & 21.9(0.5) \\
\hline
H$_2$$^{17}$O \\
\hline
$1_{11}-0_{00}$ & 1107166.9 & 53.1 & 3.44 & 102.3(10.3) & 13.7(0.5) & 23.6(0.8) \\
$1_{10}-1_{01}$ & 552021.0 & 60.7 & 15.48 & 96.4(9.7) & 11.9(0.1) & 26.7(0.2) \\
$2_{12}-1_{01}$ & 1662464.4 & 114.0 & 15.48 & 204.5(20.8) & 13.6(0.2) & 22.5(0.4) \\
$2_{11}-2_{02}$ & 748458.3 & 136.6 & 7.11 & 128(26) & 9.4(2.0) & 25.0 \\
$2_{21}-2_{12}$ & 1646398.7 & 193.0 & 8.60 & 112.4(13.3) & 15.1(0.9) & 25.0 \\
$2_{20}-2_{11}$ & 1212980.4 & 194.9 & 4.36 & 104.6(11.1) & 11.0(0.4) & 23.7(0.8) \\
$3_{03}-2_{12}$ & 1718119.5 & 196.5 & 18.08 & 125.5(15.5) & 16.3(1.1) & 25.0 \\
$3_{12}-3_{03}$ & 1096414.3 & 249.1 & 22.38 & 223.8(22.4) & 9.7(0.1) & 25.0 \\
$3_{21}-3_{12}$ & 1148976.1 & 304.2 & 26.31 & 220(22) & 10.3(0.2) & 25.0 \\
$4_{13}-4_{04}$ & 1604179.9 & 395.9 & 6.98 & 36.4(11.6) & 12.7(5.4) & 25.0 \\
$4_{22}-4_{13}$ & 1197610.3 & 453.3 & 12.54 & 67.2(25.3) & 5.5(1.5) & 17.8(6.8) \\
\hline
HDO \\
\hline
$1_{11}-0_{00}$ & 893638.7 & 42.9 & 3.0 & 120.7(12.1) & 8.4(0.1) & 23.0(0.2) \\
$1_{10}-1_{01}$ & 509292.4 & 46.8 & 4.52 & 60.4(6.7) & 9.1(0.4) & 21.0(0.9) \\
$2_{02}-1_{01}$ & 919310.9 & 66.4 & 0.86 & 90.2(9.0) & 9.6(0.1) & 20.6(0.1) \\
$2_{02}-1_{11}$ & 490596.6 & 66.4 & 1.91 & 62.0(6.2) & 8.6(0.1) & 18.6(0.1) \\
$2_{12}-1_{01}$ & 1277675.9 & 83.6 & 4.53 & 99.7(10.0) & 10.1(0.2) & 23.2(0.2) \\
$2_{12}-1_{11}$ & 848961.8 & 83.6 & 0.65 & 70.0(7.1) & 8.0(0.1) & 18.0(0.3) \\
$2_{11}-1_{10}$ & 1009944.7 & 95.2 & 0.65 & 74.1(10.1) & 9.0 & 19.0 \\
$2_{11}-2_{02}$ & 599926.7 & 95.2 & 6.87 & 53.2(5.4) & 9.2(0.5) & 19.3(0.5) \\
$3_{03}-2_{12}$ & 995411.5 & 131.4 & 4.30 & 110.0(11.0) & 8.6(0.1) & 21.8(0.2) \\
$3_{13}-2_{02}$ & 1625408.1 & 144.4 & 6.29 & 51.9(11.2) & 9.0 & 20.0 \\
$2_{21}-2_{12}$ & 1522925.8 & 156.7 & 2.51 & 61.3(7.1) & 9.0 & 20.0 \\
$3_{12}-2_{11}$ & 1507261.0 & 167.6 & 1.16 & 44.1(6.0) & 9.0 & 20.0 \\
$3_{12}-3_{03}$ & 753411.2 & 167.6 & 8.30 & 104.7(11.3) & 9.2(0.4) & 22.4(0.9) \\
$4_{04}-3_{13}$ & 1491926.9 & 216.0 & 7.15 & 72.7(8.8) & 11.2(0.9) & 25.9(1.5) \\
$4_{14}-3_{13}$ & 1678577.8 & 225.0 & 1.62 & 25.3(6.6) & 11.9(2.9) & 25.0 \\
$3_{21}-3_{12}$ & 1217258.3 & 226.0 & 6.33 & 77.0(10.5) & 9.5(0.7) & 22.5(1.1) \\
$4_{13}-3_{22}$ & 827263.4 & 263.3 & 1.75 & 19.7(2.1) & 5.8(0.4) & 16.9(0.8) \\
$4_{23}-4_{14}$ & 1818529.7 & 312.3 & 5.30 & 27.8(7.9) & 10.5(1.9) & 21.0 \\
$4_{22}-4_{13}$ & 1164769.9 & 319.2 & 9.81 & 79.7(8.3) & 7.4(0.2) & 22.0(0.6) \\
$6_{24}-6_{15}$ & 1230402.9 & 580.6 & 15.82 & 40.4(4.6) & 7.5(0.3) & 21.0
\enddata
\tablenotetext{a}{Numbers without uncertainties indicate values that were not varied in the fit.}
\end{deluxetable}

\clearpage
\begin{deluxetable}{c c c c c c c c}
\tablewidth{0pt}
\tablecaption{Fit integrated fluxes for H$_2$O isotopologues in the absorbing gas.}
\tablehead{
Transition & Frequency & $E_l$ & $S_\textnormal{ij}\mu^2$ & $\Delta T_\textnormal{abs}$ & $|\Delta T_\textnormal{abs}/T_\textnormal{bg}|$ & $v_\textnormal{LSR}$\tablenotemark{a} & $\Delta v$\tablenotemark{a} \\
 & (MHz) & (K) & (D$^2$) & (K) & &  (km s$^{-1}$) & (km s$^{-1}$)}
\startdata
H$_2$$^{18}$O \\
\hline
$1_{11}-0_{00}$ & 1101698.3 & 0.0 & 3.44 & -2.9(0.6) & 0.32(0.07) & -5.1 & 30.0 \\
$2_{12}-1_{01}$ & 1655867.6 & 34.2 & 15.49 & -7.3(0.8) & 0.44(0.05) & -5.1 & 30.0 \\
$2_{21}-2_{12}$ & 1633483.6 & 113.7 & 8.60 & -1.52(0.25) & 0.092(0.015) & -5.1 & 30.0 \\
$3_{03}-2_{12}$ & 1719250.2 & 113.7 & 18.16 & -2.94(0.37) & 0.172(0.022) & -5.1 & 30.0 \\
\hline
H$_2$$^{17}$O \\
\hline
$1_{11}-0_{00}$ & 1107166.9 & 0.0 & 3.44 & -1.59(0.19) & 0.171(0.021) & -5.1 & 30.0 \\
$2_{12}-1_{01}$ & 1662464.4 & 34.2 & 15.48 & -3.33(0.44) & 0.202(0.026) & -5.1 & 30.0 \\
\hline
HDO \\
\hline
$1_{11}-0_{00}$ & 893638.7 & 0.0 & 3.02 & -0.44(0.11) & 0.081(0.020) & -5.1 & 25.0 \\
$2_{12}-1_{01}$ & 1277675.9 & 22.3 & 4.53 & -1.36(0.14) & 0.103(0.011) & -5.1 & 25.0
\enddata
\tablenotetext{a}{Numbers without uncertainties indicate values that were not varied in the fit.}
\end{deluxetable}

\end{document}